\documentclass[12pt,a4paper]{article}

\usepackage[a4paper,top=2cm,bottom=2cm,left=2.5cm,right=2.5cm,marginparwidth=1.75cm]{geometry}
\usepackage[sorting=none]{biblatex}
\usepackage{amsmath}
\usepackage{amssymb}
\usepackage{braket}
\usepackage{amsfonts}
\usepackage{graphicx}
\usepackage{hyperref}
\usepackage{authblk}

\title{Second-Order Raman Scattering in Exfoliated Black Phosphorus}
\author[1]{Alexandre Favron}
\author[1]{F\'elix Antoine Goudreault}
\author[1]{Vincent Gosselin}
\author[1]{Julien Groulx}
\author[1]{Michel C\^ot\'e}
\author[1]{Richard Leonelli}
\author[2]{Jean-Francis Germain}
\author[2]{Anne-Laurence Phaneuf-L'Heureux}
\author[2]{S\'ebastien Francoeur}
\author[3]{Richard Martel}
\affil[1]{D\'{e}partement de physique and 
          Regroupement Qu\'{e}b\'{e}cois sur les Mat\'{e}riaux de Pointe, Universit\'{e} de Montr\'{e}al, C.~P. 6128, 
          Succursale Centre-Ville, Montr\'{e}al, Qu\'{e}bec H3C~3J7, Canada.}
\affil[2]{D\'{e}partement de g\'{e}nie physique, 
          \'{E}cole Polytechnique de Montr\'{e}al, C.~P. 6079,
          Succursale Centre-ville, Montr\'{e}al, Qu\'{e}bec H3C~3A7, Canada
          }
\affil[3]{D\'{e}partement de chimie and 
          Regroupement Qu\'{e}b\'{e}cois sur les Mat\'{e}riaux de Pointe, Universit\'{e} de Montr\'{e}al,
          C.~P. 6128, Succursale Centre-Ville, Montr\'{e}al, Qu\'{e}bec H3C~3J7, Canada}
\affil[*]{\texttt{michel.cote@umontreal.ca}}

\begin{document}

\maketitle





\begin{abstract}
Second-order Raman scattering has been extensively studied in carbon-based nanomaterials, \emph{e.g.} nanotube and graphene, because it activates normally forbidden Raman modes that are sensitive to crystal disorder, such as defects, dopants, strain, etc. The sp$^2$-hybridized carbon systems are, however, the exception among most nanomaterials, where first-order Raman processes usually dominate. Here we report the identification of four second-order Raman modes, named $D_1$, $D_1'$, $D_2$ and $D_2'$, in exfoliated black phosphorus (P(black)), an elemental direct-gap semiconductor exhibiting strong mechanical and electronic anisotropies. Located in close proximity to the $A^1_g$ and $A^2_g$ modes, these new modes dominate at an excitation wavelength of 633 nm. Their evolutions as a function of sample thickness, excitation wavelength, and defect density indicate that they are defect-activated and involve high-momentum phonons in a doubly-resonant Raman process. \emph{Ab initio} simulations of a monolayer reveal that the $D'$ and $D$ modes occur through intravalley scatterings with split contributions in the armchair and zigzag directions, respectively. The high sensitivity of these $D$ modes to disorder helps explaining several discrepancies found in the literature.
\end{abstract}

\section{\label{sec:intro}Introduction}

Raman spectroscopy produces rich signals involving not only first-order Raman scattering (one phonon processes), but also higher-order processes that can be efficiently exploited to characterize 2D materials.~\cite{Ferrari2013,Zhang15} An intriguing example is the doubly-resonant Raman (DRR) mode~\cite{Carvalho2013}, known also as the phonon-defect mode, in which an electron-phonon interaction is activated by an impurity, a localized defect, or an edge, and involves a phonon with large quasi-momentum, $q$. As prime examples of the phonon-defect modes ($D$ modes), the so-called $D$ and $D'$ bands in graphene have played an instrumental role in the development of graphene because they provided valuable information on the number of layers~\cite{ferrari_raman_2006}, the carrier mobility~\cite{Ferrari2013}, the level of doping~\cite{Das2008}, the presence of disorder~\cite{Lucchese2010}, and  phonon dispersion~\cite{mafra_determination_2007}. Under specific wavelengths excitation, few studies only have reported the presence of phonon-defect modes in other 2D materials and most, if not all, involved dichalcogenides, such as MoS$_2$~\cite{Carvalho2013,Mignuzzi2015,Wu2016}, WS$_2$~\cite{Tan2017}, WSe$_2$~\cite{Shi2016} and ReS$_2$~\cite{McCreary2017}. 
Black phosphorus (P(black)), an elemental semiconductor of P atoms~\cite{Brown65,Keyes53}, is a particularly interesting case since it presents some of the key elements found in graphene. Besides its lamellar structure, band structure calculations of the isolated monolayer of P(black) predict low energy bands and multiple valleys~\cite{Cai2014}, which can promote, in given conditions, high-order resonances. More specifically, the electronic band structure of the monolayer exhibits a single valence band at the center of the Brillouin zone and two valleys in the conduction band, one of lowest energy centered at $\Gamma$ and another in the zigzag direction ~\cite{Cai2014}.
Compared to graphene, few-layer P(black) is, however, significantly different because a reduction of the layer thickness increases the bandgap from 0.34 eV for the bulk~\cite{Morita86} to $\sim$1.7 eV for the monolayer~\cite{yang_optical_2015,Li2016gap}, whereas graphene remains semimetallic irrespective of the number of layers. Furthermore, P(black) material is anisotropic~\cite{Qiao2014}, which makes it particularly interesting to explore high-order resonances using polarization-resolved Raman spectroscopy. 

The monolayer of P(black) is often called phosphorene in the recent literature, but the name 2D-phosphane is used hereafter to better comply with IUPAC nomenclature.~\cite{favron_photooxidation_2015} Of the six Raman-active modes of the bulk P(black), only three are allowed in the backscattering configuration for all thicknesses down to the monolayer: $A^1_g$, $B_{2g}$ and $A^2_g$ ~\cite{favron_photooxidation_2015,wang_highly_2015,lu_plasma-assisted_2014,liu_phosphorene_2014}. In Ref.~\cite{phaneuf-lheureux_polarization-resolved_2016}, we fully characterized these modes for thin layers with polarization-resolved Raman spectroscopy and reported a new first-order mode, labeled $A^2_g(B_{2u})$, in samples with a number of atomic layers, $n$, between $2$ and $5$. The latter is nearly degenerate with $A^2_g$ and was assigned to a Davydov-induced conversion of the $B_{2u}$ infrared (IR) mode. Additional unexpected Raman features were also observed from pristine and oxidized 2D-phosphane for thicknesses ranging from $n~=~1$ to $5$~:  i) relatively intense shoulders or well-resolved peaks in the close vicinity of $A^1_g$ and $A^2_g$; ii) a wide band sensitive to degradation located between $B_{2g}$ and $A^2_g$~\cite{favron_photooxidation_2015,phaneuf-lheureux_polarization-resolved_2016}; and iii) an evolution of the $A^1_g$/$A^2_g$ integrated intensity ratio during sample oxidation in ambient conditions~\cite{favron_photooxidation_2015}. Unexpected from a first-order Raman analysis, these new Raman features have not yet been the subject of a systematic experimental study and their origins remain largely unexplained.

In this article we present a Raman study of exfoliated P(black) samples with $n$ = 1-7, 9, 12 and 18 layers performed at three excitation wavelengths ($\lambda_{ex}$~=~488~nm, 532~nm and 633~nm). The spectra are analyzed and structured according to the explicit trends observed for the mode frequency and intensity as a function of $n$ and excitation wavelengths. Combined with polarization-resolved experiments, this procedure leads to the identification of four new totally-symmetric modes, labelled $D_1$, $D_1'$ and $D_2$, $D_2'$, which are located very close to the bulk $A^1_g$ and $A^2_g$ modes, respectively. The degradation kinetics of the trilayer ($n$ = 3) 2D-phosphane indicates that these new Raman modes are sensitive to the presence of defects. \emph{Ab initio} calculations of the monolayer electronic and phonon band structures and Raman spectra simulations support the assignment that these $D$ modes are defect-phonon modes due to second-order Raman involving high-$q$ phonons with dominant contributions either along the zigzag ($D$) or armchair ($D'$) directions. This assignment is further supported by the observed dependence of the mode intensities on laser frequency and by the change in the intensity ratio of the second to the first-order $A_g$ modes with degradation time. Finally, the properties of all Raman modes are reviewed to explain various behaviors reported in the literature, such as shifts in mode frequencies with layer thickness and changes in $A^1_g$/$A^2_g$ ratio with degradation.

\section{\label{sec:results}Results}

\begin{figure*}[!hbp]
 \includegraphics[trim={0cm 0cm 0cm 0cm},clip,width=16cm]{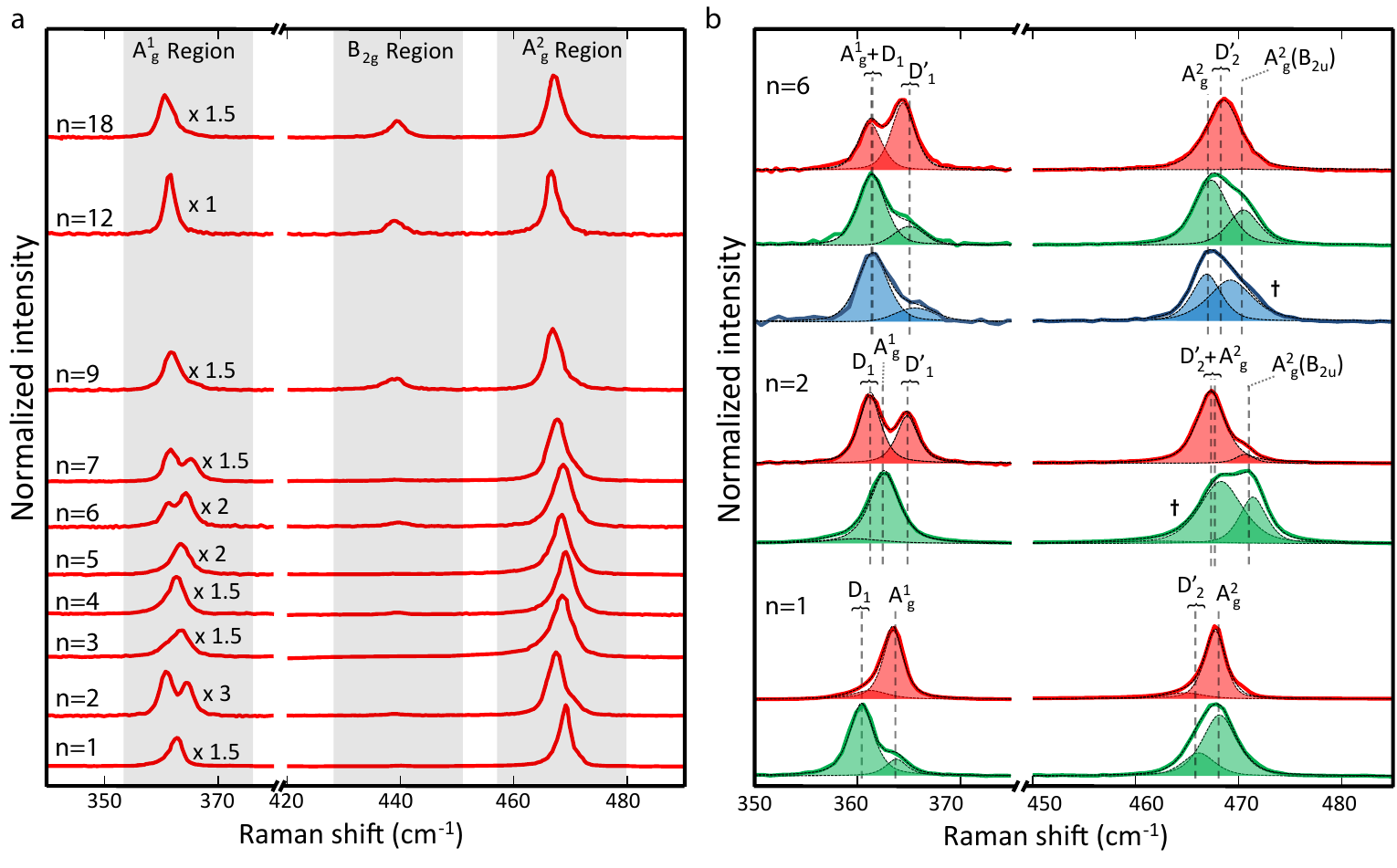}
        \caption[Raman experiments on 2D-phosphane samples with $n$~=~1-7, 9, 12
        and 18 layers.]
        {Raman experiments on 2D-phosphane samples with $n$~=~1-7, 9, 12 and 18 layers 
        deposited on a SiO$_2$/Si substrate and measured at 300 K. 
        $\bf{a}$, Raman spectra taken at $\lambda_{ex}$~=~633~nm. $\bf{b}$, 
        Zoom of the multimodal structures in the $A^1_g$ and $A^2_g$ regions at 
        $\lambda_{ex}$~=~488~nm (blue), 532~nm (green) and 633~nm (red). 
        In panel $\bf{a}$, the spectra are normalized relative to the $A^2_g$ maximum and they are vertically shifted and scaled to enhance the weaker $A^1_g$ regions. 
 Fit to the multimodal peaks is shown in panel $\bf{b}$ along with the expected position of each of the $D$ modes. The spread in frequency of the $D$ modes are roughly shown using brackets. Dagger symbols mark the wider peaks and highlight nearly degenerated modes.}
 \label{Fig1}
  \end{figure*}

Bulk P(black) has a base-centered orthorhombic Bravais lattice, a primitive cell containing 4 atoms, and a $D^{18}_{2h}$ space group symmetry~\cite{Sugai85}. Although this group yields 6 Raman-allowed modes, the out-of-plane excitation used in typical backscattering experiments allows for the observation of only three modes: $A^1_g$, $A^2_g$ and $B_{2g}$. The $A_g$ Raman tensor is composed of three distinct diagonal elements while that of $B_{2g}$ is composed of two identical non-diagonal elements. 
For 2D-phosphane, the factor-group $D_{2h}$ is preserved irrespective of the sample thickness, which implies that all vibrational modes are non-degenerate and that the bulk nomenclature for identifying these modes remains valid for all $n$. Fig.~\ref{Fig1}a presents typical Raman spectra measured at 300~K from several $n$-layer 2D-phosphane samples with an excitation wavelength of $\lambda_{ex}$~=~633~nm and in the polarization configuration which maximizes the $A_g$ intensity ($\theta_{ex.}$ and $\theta_{meas.}$ both aligned along the armchair direction). 
As $B_{2g}$ did not reveal any behavior not already reported in the literature, the following  discussion is limited to the totally-symmetry modes. In contrast with the single vibrational mode expected from a first-order Raman analysis, excitation at 633 nm leads to rather complex multimode structures in the close vicinity of both $A^1_g$ and $A^2_g$. The spectra of Fig.~\ref{Fig1}a alone do not allow resolving the overlapping modes, but, since their relative intensities are sensitive to the excitation wavelength, an analysis of the Raman responses obtained at three distinct excitation wavelengths (488, 532, and 633 nm) allows discriminating the various contributions to these multimode structures (see Fig.~\ref{Fig1}b and supplementary, Fig.~S1). For example, the bilayer multimode structure in the vicinity of $A_g^1$ is the sum of three contributions at distinct frequencies: a single mode dominating at 532 nm and two additional and clearly resolved modes at 633 nm. 
  
\begin{figure*}[!htp]
 \includegraphics[trim={0cm 0cm 0cm 0cm},clip,width=16cm]{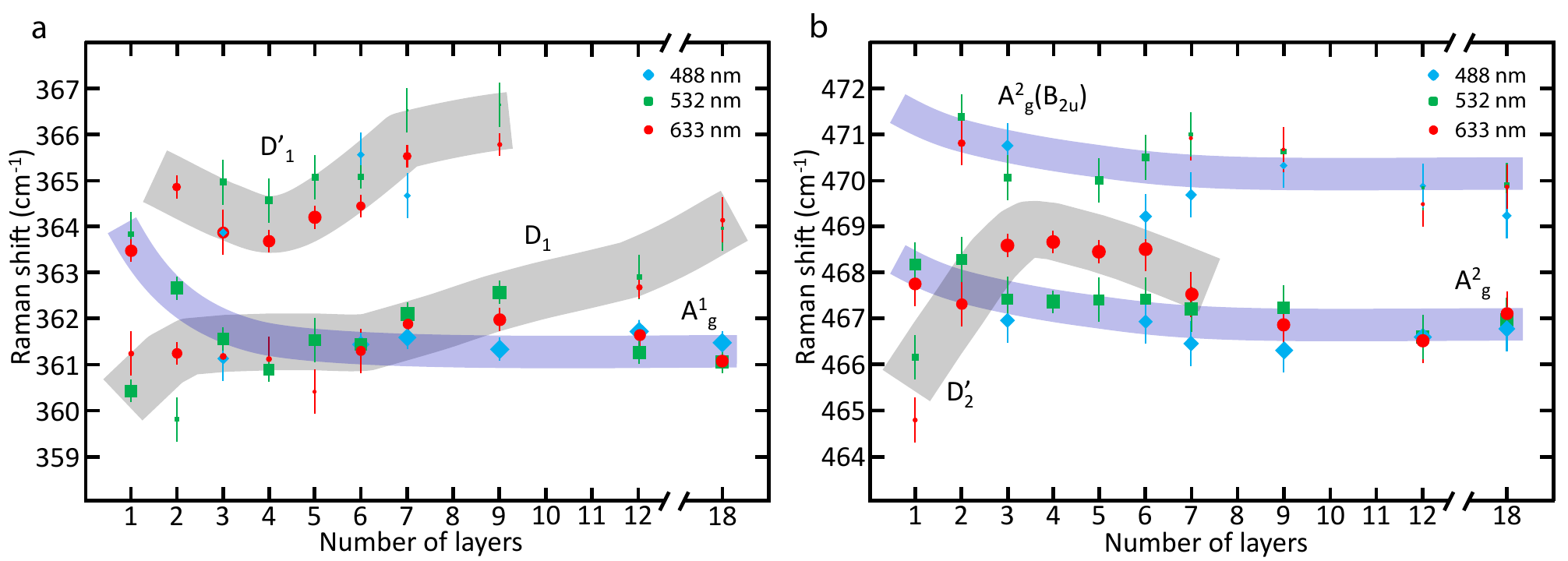}
        \caption[Evolution of the first-order and second-order Raman peaks as a
        function of the number of layers and wavelength excitation in the
        $A_g^1$ and $A^2_g$ regions.]
        {Evolution of the first-order (blue-stripes) and second-order (gray stripes) Raman 
        peaks as a function of the number of layers, $n$, and wavelength 
        excitation in the $A^1_g$ ($\bf{a}$) and $A^2_g$ ($\bf{b}$) regions. The area of the markers is proportional to the integrated intensity relative to each region. The blue stripes are guides to the eyes for the assignations of the $A^1_g$, $A^2_g$ and $A^2_g(B_{2u})$ modes and the gray stripes are the same for the $D_1$, $D_1'$ and  $D_2'$ modes. The widths of the stripes indicate approximately the spread of the $D$ modes.}
  \label{Fig2}
\end{figure*}

The frequency and intensity of all Raman modes identified in Figs.~\ref{Fig1} and ~S1 are reported in Fig.~\ref{Fig2} as a function of the number of layer, $n$, and excitation wavelength. These results suggests the presence of 6 Raman modes in these two narrow frequency ranges: $D_1$ and $D_1'$ are found near $A_g^1$ and  $D_2'$ accompanies $A_g^2$ and  $A_g^2(B_{2u})$. Several important considerations were used to discriminate the overlapping evolution of these Raman modes with thickness. First, the spectra measured on thick samples ($n>10$) should be dominated by the well-established first-order bulk P(black) vibrational modes and should not be sensitive to excitation wavelength. Second, the frequency evolution of first-order allowed $A_g^2$ and Davydov-induced $A_g^2(B_{2u})$ with thickness should be very similar, since they involve nearly degenerate but out-of-phase atomic motions in the armchair direction~\cite{phaneuf-lheureux_polarization-resolved_2016} and have been shown to easily couple and mix character~\cite{Sun2016}. Third, the intensity of $D$ and $D'$ modes generally increases with the excitation wavelength, reaching a maximum at 633~nm and making it easier to discriminate between first-order ($A_g^1$, $A_g^2$ and  $A_g^2(B_{2u})$) and $D$ or $D'$ modes. The forth consideration involves resonant effects in the mono- and bilayer. At $\lambda_{ex}$~=~633~nm (1.96 eV), the excitation is in near resonance with the optical bandgap of the monolayer ($\sim$1.75 eV)~\cite{yang_optical_2015,Li2016gap}. Similarly, excitation at $\lambda_{ex}$~=~532~nm (2.33 eV) is resonant with the transition between the second valence and second conduction bands of the bilayer. \cite{favron_photooxidation_2015,phaneuf-lheureux_polarization-resolved_2016,Li2016gap} These resonant conditions singularly enhance the Raman response from first-order modes and help identify the frequencies of both $A^1_g$ and $A^2_g$ in the monolayer and bilayer. Finally, polarization-resolved Raman experiments on the D modes reveal that the atomic motions involve crystalline vibrations associated to the $A_g$ representation. As a representative example, the polar dependence of $D_1'$ on a trilayer in a parallel configuration ($\theta_{ex.}=\theta_{meas.}$), shown in Fig.~S3f (supplementary), is characteristics of a diagonal Raman tensor composed of two anisotropic elements ($R_{aa}$ and $R_{cc}$). Like the case for bulk $A_g$ modes, the armchair element dominates ($R_{cc}>R_{aa}$) at lambda=532 nm.~\cite{phaneuf-lheureux_polarization-resolved_2016}.

These considerations provides the rationale for the mode assignment illustrated by the colored bands in Fig.~\ref{Fig2}: the purple and gray stripes represent the evolution with thickness of first-order ($A_g^1$, $A_g^2$ and $A_g^2(B_{2u})$) and D ($D_1$, $D_1'$, and $D_2'$) modes,  respectively. All first-order vibrational modes soften with increasing layer thickness. This observed trend matches that calculated from density functional theory~\cite{Hu2016} and confirms the overall validity of the mode assignment. Both $D_1'$ and $D_2$ increase in frequency with $n$, but $D_2'$ appears to exhibit a non-monotonic dependence. The intensity of $D$ modes fades with $n$, which is consistent with bulk-forbidden modes activated through a mechanism that is not of first order. As will be discussed later, no simple thickness dependence can be expected from $D$ modes. 

\begin{figure*}
 \includegraphics[trim={0cm 0cm 0cm 0cm},clip,width=17cm]{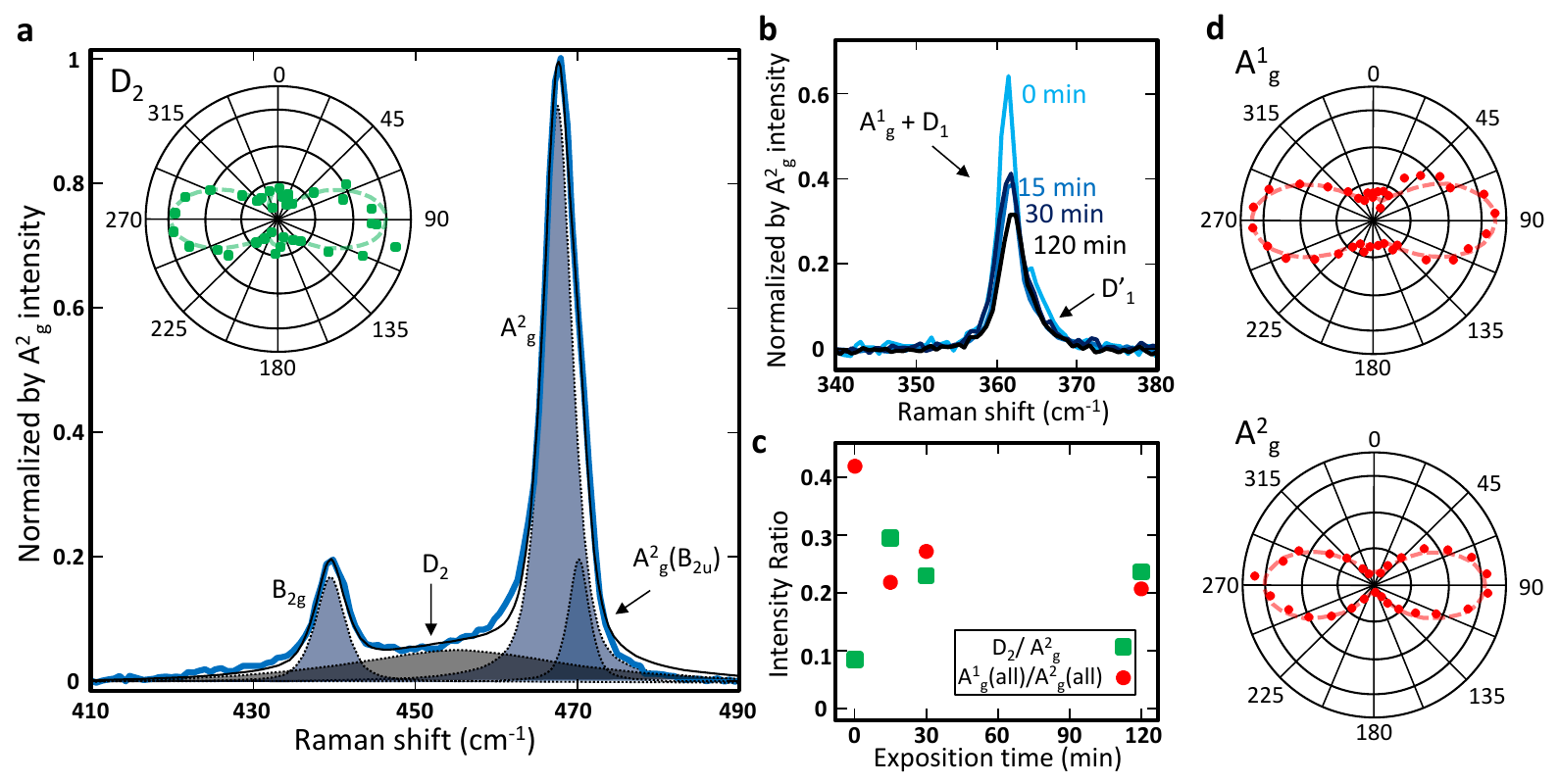}
        \caption[Presentation of the mode $D_2$ along with its polarization and
        evolution during degradation in ambient conditions.]
        {Presentation of the mode $D_2$ along with its polarization properties and 
        evolution during degradation in ambient conditions. $\bf{a}$, Sample with $n$~=~5 layers at $\lambda_{ex}$~=~532~nm after an exposure of 15 minutes to ambient conditions at T = 300 K. Inset: a polarization plot of the Raman intensity of the $D_2$ mode in the parallel  configuration ($\theta_{exc}$=$\theta_{mes}$) after 120 minutes in ambient conditions. $\bf{b}$, Spectra of the $A^1_g$ region normalized with the $A^2_g$ peak after 0, 15, 30, and 120 min exposures in ambient conditions. $\bf{c}$, Temporal evolution of integrated intensity of the $D_2$/$A^2_g$ (green square) and $A^1_g$/$A^2_g$ (red circle) ratios. $\bf{d}$, Polarization plot of the Raman intensity of the $A^1_g$ and $A^2_g$ modes in the parallel configuration after 120 minutes in ambient conditions. The spectra in $\bf{a}$ is the sum of all of the parallel polarization spectra.}
  \label{Fig3}
\end{figure*}

  The results presented above were obtained from pristine samples exfoliated in a controlled environment (See \hyperref[sec:method]{Methods}). However, important clues on the origin of $D$ modes can be obtained from samples exposed to conditions known to induced degradation~\cite{Wood14,favron_photooxidation_2015}. We discuss next the appearance of a fourth $D$ mode and the change in intensity of $D_1$, $D_1'$, and $D_2'$, which are found to be much more sensitive to degradation than first-order modes. These observations suggest that additional scattering mechanisms are activated by defects.   
  
   $D_2$ appears when pristine black phosphorus is exposed to ambient light and humidity conditions. As shown in Fig.~\ref{Fig3}a, a broad band appears underneath $B_{2g}$ and $A^2_g$ after exposing this 5-layer 2D-phosphane sample to ambient light and atmospheric conditions for 15 minutes. This feature can often be observed in samples of suboptimal quality~\cite{favron_photooxidation_2015}, but it has never been discussed. Subsequent exposures to ambient conditions lead to the spectra shown in Fig.~S2. The ratio of $D_2$ intensity to that of $A_g^2$ is shown by the green squares in Fig.~\ref{Fig3}c and shows that the $D_2$ relative intensity grows for the first 15 min of exposure and then remains constant for up to 120 min of cumulative exposure. The polarization response from $A^1_g$, $A^2_g$ and $D_2$ after 120 min are shown in Fig.~\ref{Fig3}a,d. As for the other D modes, the polarization dependence of $D_2$ is that of a totally symmetric vibrational mode. An analysis of the Raman tensor elements as a function of exposure time is presented in supplementary, Fig.~S3. Note that the polarization-resolved Raman intensities were adjusted with the dichroism of $n$ = 3 samples and the birefringence of black phosphorus, following the procedure described elsewhere~\cite{phaneuf-lheureux_polarization-resolved_2016}. A detailed analysis of $D_2$ is complicated by its large linewidth, hence we focus next on the other three $D$ modes first presented in Fig.~\ref{Fig2}. 
      
\begin{figure*}[!hbp]
 \includegraphics[trim={0cm 0cm 0cm 0cm},clip,width=14cm]{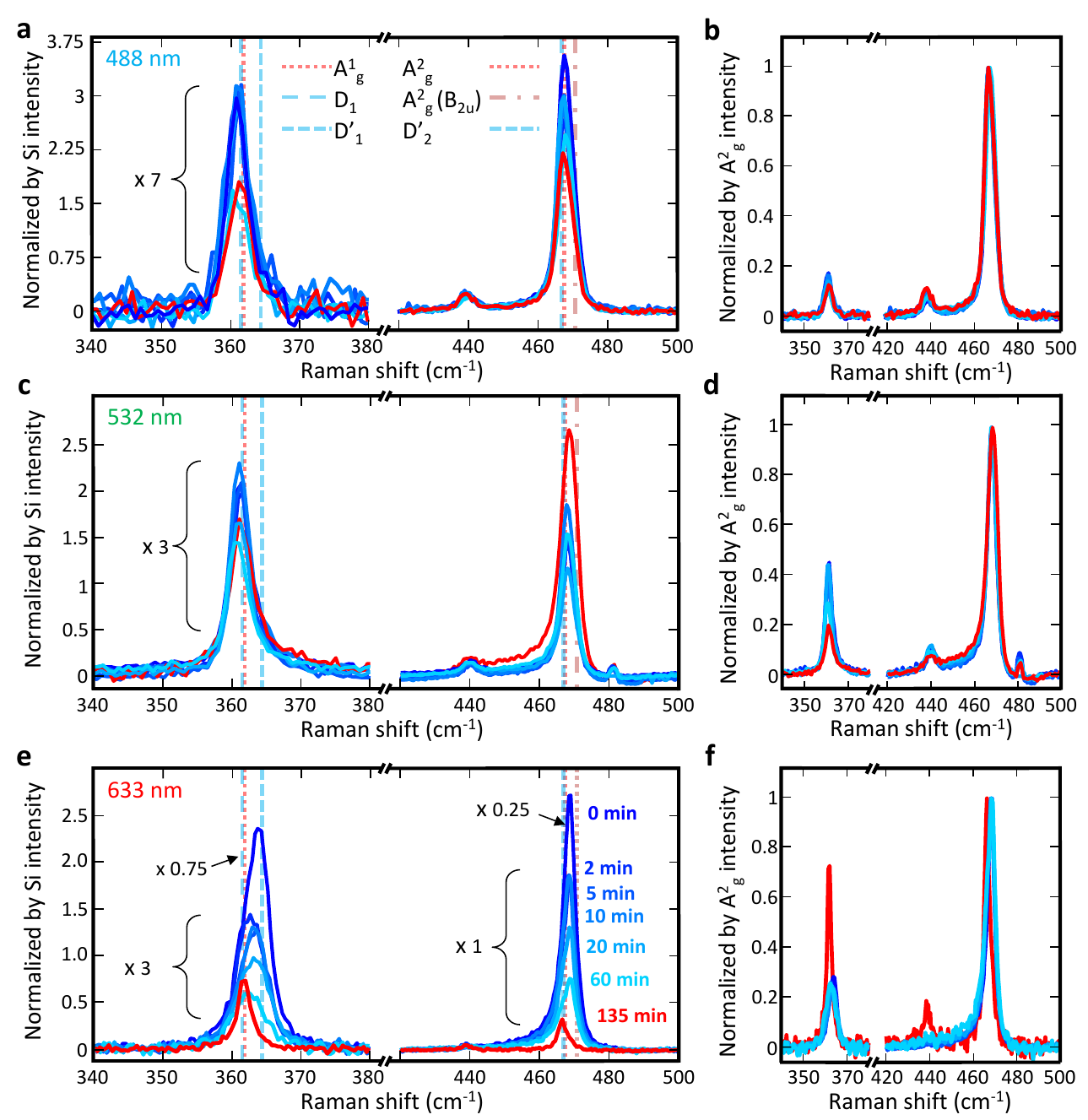}
        \caption[Degradation experiment in ambient conditions on a trilayer
        2D-phosphane.]
        {Degradation experiment in ambient conditions on a trilayer 2D-phosphane 
        at $\lambda_{ex}$~=~488~nm ($\bf{a,b}$), 532~nm ($\bf{c,d}$) and 633~nm ($\bf{e,f}$) 
        under a constant fluence of 400 $\mu W/\mu m^2$ . The $A^1_g$ and $A^2_g$ regions in panels $\bf{a}$, $\bf{c}$ and $\bf{e}$ are normalized to the silicon peak at $\sim$520 cm$^{-1}$. Panels $\bf{b}$, $\bf{d}$ and $\bf{f}$ present the complete spectra normalized by the $A^2_g$ intensity.}
  \label{Fig4}
\end{figure*}

At 532 nm, the integrated intensity ratio  of all Raman features in the vicinity of $A_g^1$ to that of all those in the vicinity of $A_g^2$, \emph{i.e.} the $A_g^1(\text{all})/A_g^2(\text{all})$ ratio, decreases with exposure time to oxidation in ambient conditions (red circles in Fig.~\ref{Fig3}c), as previously reported~\cite{favron_photooxidation_2015, Hu2017}. This ratio can be used as a marker for sample quality, but its strong variation is somewhat unexpected from first-order contributions alone, suggesting again that defect-activated scattering mechanisms are involved. This is further illustrated in Fig.~\ref{Fig4}, where the absolute Raman signal measured from a trilayer 2D-phosphane is shown as a function of exposure time to ambient light and humidity conditions and measured at three distinct excitation wavelengths, 488~nm (panels $a$,$b$), 532~nm (panels $c$,$d$) and 633~nm (panels $e$,$f$). To emphasize the effect of degradation on relative intensities, spectra normalized to the intensity of $A^2_g$ mode are shown in panels $b$, $d$, and $f$. As noted earlier, the intensity of $D$ modes are enhanced at longer excitation wavelengths. Therefore, the degradation dynamics at 488 nm (panel a and b) are minimally influenced by $D$ modes, which are overlapping with $A_g^1$ and $A_g^2$ in the trilayer, and mostly determined by first-order contributions. Although a net decrease in intensity is observed (panel $a$), the relative intensity of $A_g^1$ (panel $b$) barely evolves up to an exposure time of 135 min, indicating that $A_g^1$ and $A_g^2$ are similarly affected by sample degradation. At 532 nm, the contribution from $D$ modes becomes more important and significantly affects the relative intensity of $A_g^1$. The intensity evolution observed in Fig.~\ref{Fig3}c can be explained by the sensitivity of $D$ modes to degradation and the fact that two $D$ modes are found near $A_g^1$ against only one near for $A_g^2$. 

As already illustrated in Fig.~\ref{Fig2}, $D_1'$ and $D_2'$ dominate over first-order modes at an excitation wavelength of 633 nm (Fig.~\ref{Fig4}e). However, degradation quickly quenches their intensity, resulting in a significant red-shift and narrowing of the Raman lines. After 135 min, the spectrum is dominated by first-order modes $A_g^1$ and $A_g^2$. Although the intensity of both $A_g^1$ and $A_g^2$ is affected by degradation, their relative ratio is not (panel $f$), confirming that bulk-like modes are equally affected by degradation.  

Our Raman results unambiguously identify several new bulk-forbidden modes in few-layers 2D-phosphane: $D_1$, $D_1'$, $D_2$ and $D_2'$. 
These features add to the other bulk-forbidden modes discussed so far in the literature for $n\geq2$ samples: i) The Davydov-induced $A^2_g(B_{2u})$ mode~\cite{phaneuf-lheureux_polarization-resolved_2016}, already presented above, and ii) the breathing modes identified in the low frequency region of the spectrum ($\leq$~$100$~cm$^{-1}$)~\cite{Ling2015,Dong16}. These modes involve first-order Raman scattering, whereas the present study suggests that the $D$ modes are associated to higher-order Raman scattering, a mechanism that is not obvious at first sight. Some elements come, however, in support of a second-order mechanism involving the scattering of a large-$q$ phonon and a defect. First, the $D$ modes are present even in the case of the monolayer, which rules out first-order processes such as Davydov-induced modes due to interlayer coupling, Davydov broadening, and other mechanisms related to interlayer interactions. Second, the $A_g$ symmetry of $D_1'$ and $D_2$, as determined by polarization experiments in the inset of Fig.~\ref{Fig3}a (supplementary, Fig.~S3), indicates that they are crystalline vibrations. Third, the degradation experiments in Fig.~\ref{Fig3}a,c demonstrate that the presence of defects increases the relative intensity of the $D_2$ mode. Combined together with the analysis in Fig.~\ref{Fig4}, it appears that all four $D$ modes share a similar origin. 

\begin{figure*}[!tbp]
 \includegraphics[trim={0cm 0cm 0cm 0cm},clip,width=17cm]{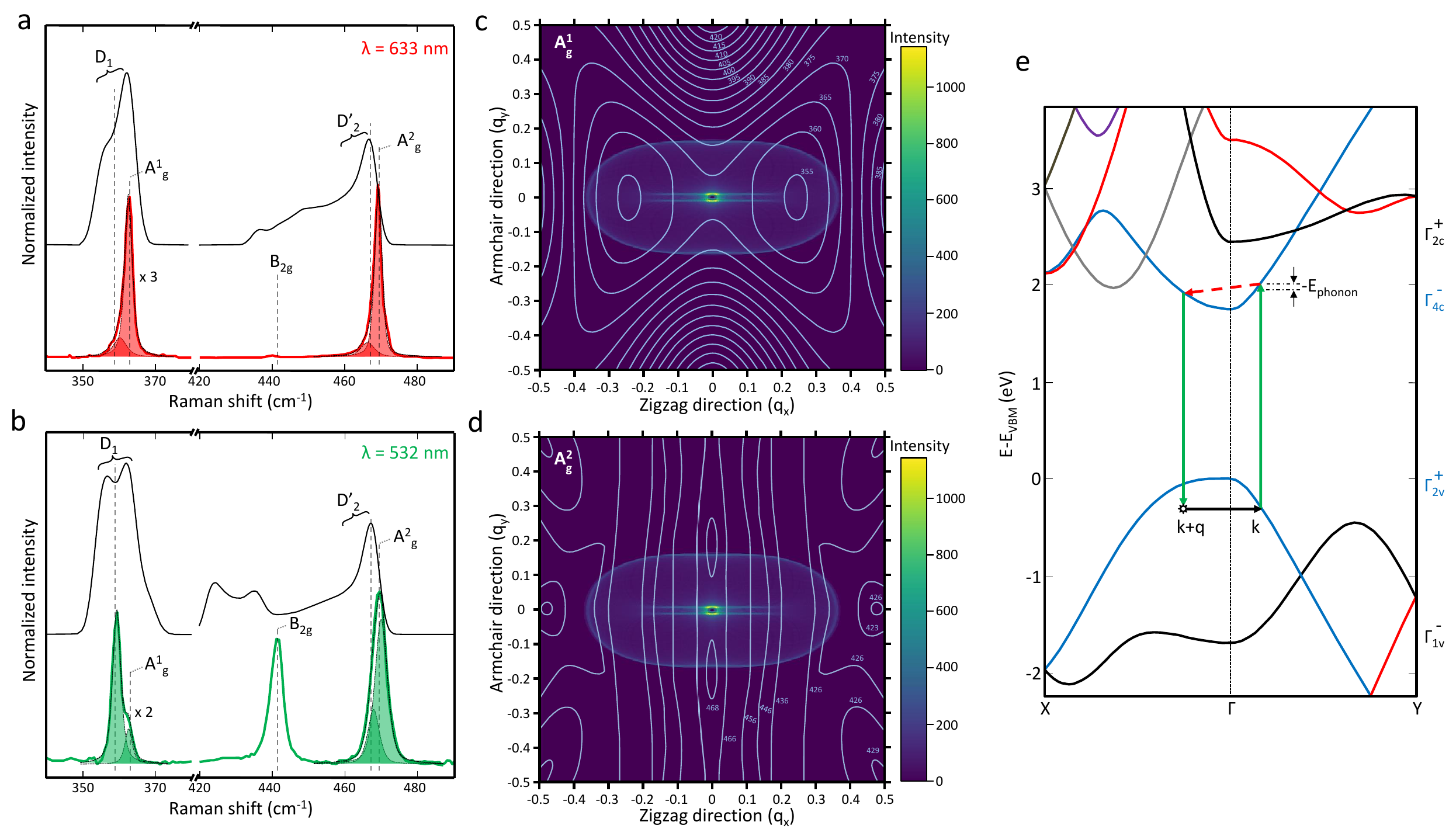}
        \caption[Phonon-defect scattering model in a monolayer 2D-phosphane.]
        {Phonon-defect scattering model in a monolayer 2D-phosphane. $\bf{a}$, 
        Calculated phonon-defect modes of a monolayer (black curve) and corresponding Raman spectrum at $\lambda_{ex}$~=~633~nm (red curve). $\bf{b}$, Calculated phonon-defect modes of a monolayer (black curve) and corresponding Raman spectrum at $\lambda_{ex}$~=~532~nm (green curve). $\bf{c,d}$, Momentum histogram of the phonon involved in the phonon-defect process at $\lambda_{ex}$~=~532~nm superposed on top of the calculated phonon dispersion for the $A^1_g$ and $A^2_g$ modes, respectively. $\bf{e}$, The calculated band structure for the monolayer and the symmetry character (on the right) of the bands at $\Gamma$. A sketch of an example of a phonon-defect process on top of the electronic band structure of the monolayer in which the excited electron is scattered by a phonon and the hole is scattered by a defect ($eh$). A photon is absorbed (upward arrow in green), which promotes the electron-hole pair in the conduction and valence bands. Here, the sketch depicts that the double-resonance occurs through the scattering of a high $-\boldsymbol{q}$ phonon (red dash line) in which the electron scatters to a non-virtual state with momentum $\boldsymbol{k} + \boldsymbol{q}$. The hole is scattered by a defect at $\boldsymbol{k}+\boldsymbol{q}$ (black line) and a photon is emitted when the electron recombines with the hole (downward arrow in green).}
  \label{Fig5}
\end{figure*}

Inspired by the second-order Raman effects in graphene, namely the phonon-defect ($D$ band) and the two-phonons ($2D$ band)~\cite{Ferrari2013}, we have explored theoretically all possible mechanisms for $D$ modes in 2D-phosphane. Fig.~\ref{Fig5}e presents the calculated electronic band structure (see \hyperref[sec:method]{Method}) of a monolayer 2D-phosphane, on which we have schematically drawn a phonon-defect scattering process involving a large-$q$ phonon with a localized defect. The characters of the bands is also indicated, which shows that intra-valley transitions (\emph{e.g.} between the bands in blue) are active, while inter-valley transitions (mixed coloured bands) are not possible within the energy window considered~\cite{Li2014}. As for the possible phonons, we have considered processes of both single phonon of momentum $q$ and two phonons of opposite $q$ and appropriate symmetry. Because the two-phonon condition is matched by a rather limited number of available phonons, we have calculated all of the most relevant cases, i.e. the $LA$+$LA$, $LA$+$B_{2u}$, $B_{2u}$+$B_{2u}$, $B_{1g}$+$B_{1g}$ and $B_{3g}$+$B_{3g}$ processes, and found no satisfactory combinations matching the energies of the  $A^1_g$ and $A^2_g$ regions (supplementary, section~S4). We have also considered other multi-phonon processes~\cite{Terrones2014,Ferrari2013,Thripu2014}, identified in bulk P(black) crystals using Raman~\cite{Lannin79} and mid-infrared~\cite{Ikezawa83} spectroscopies, but they have also been ruled out. That is, the reported broad responses~\cite{Morita86} are inconsistent with the sharp signatures of $D$ modes and the most relevant ones with $A_g$ symmetry, \emph{i.e.} the $B_{1g}+B_{1g}$ at 386 $cm^{-1}$,  the $2A_u-A^2_g$ at 353 $cm^{-1}$, and the $B_{3g}+B_{3g}$ at 444 $cm^{-1}$~\cite{Hu2016,Sugai85,Kaneta82,Morita83,Fei14,Cai15}, are all off in frequency. The other alternatives, \emph{e.g.} the combination of a regular mode and an acoustic phonon of small $q$, would add an asymmetrical broadening to first-order modes, which is inconsistent with our data. Furthermore, the thickness dependence involves a crossing in frequency between $A^1_g$ and $D_1$ and between $A^2_g$ and $D_2'$, which is not compatible with the two phonons hypothesis. Considering the calculated phonon diagram (supplementary, Fig.~S4a), this analysis suggests that the phonons related to the $A^1_g$ and $A^2_g$ phonon bands located in a region near the center of the zone are the most likely candidates for D modes.

The phonon-defect scattering mechanism is simulated using material properties obtained from density functional theory calculations. To highlight second-order processes, we excluded from the simulation first-order Raman scattering processes and concentrated on the second-order terms that involve one phonon and one defect. In the context of the rigid band approximation, where excited states are given by the differences in the single-electron eigenenergies, the intensity of the second-order Raman scattering mentioned above is given by the following formula~\cite{venezuela_theory_2011}:
\begin{equation}
\label{eq:drequation}
I_{\text{DR}}(\epsilon)\propto\sum_{\boldsymbol{q}, \mu}\left|
    \sum_{\boldsymbol{k},\alpha, v, c}K_\alpha(\boldsymbol{k}, \boldsymbol{q}, \mu)\right|^2
    \delta(\epsilon_L-\epsilon - \hbar\omega^\mu_{-\boldsymbol{q}})[n(\omega_{-\boldsymbol{q}}^\mu)+1]
\end{equation}
where the $K_\alpha$ represent the different possible processes labeled $\alpha$. For instance, the process, illustrated in Fig.~\ref{Fig5}e, where the excited electron is scattered by a phonon and the hole by a defect ($eh$) has an amplitude given by:   

\begin{equation}
\label{eq:Keh}
K_{eh}(\boldsymbol{k},\boldsymbol{q},\mu) = \frac{\braket{\boldsymbol{k},v|H_D|\boldsymbol{k}+\boldsymbol{q},v}
                                    \braket{\boldsymbol{k} + \boldsymbol{q},v|D_{\text{out}}|\boldsymbol{k} + \boldsymbol{q},c}
                                    \braket{\boldsymbol{k}+\boldsymbol{q},c|\Delta H_{\boldsymbol{q}\mu}|\boldsymbol{k},c}
                                    \braket{\boldsymbol{k},c|D_{\text{in}}|\boldsymbol{k},v}}
                                   {\left(\epsilon_L-\epsilon_{\boldsymbol{k}+\boldsymbol{q}}^{c}+\epsilon_{\boldsymbol{k}+\boldsymbol{q}}^v
                                   -\hbar\omega_{-\boldsymbol{q}}^\mu\right)
                                   \left(\epsilon_L-\epsilon_{\boldsymbol{k}+\boldsymbol{q}}^{c}+\epsilon_{\boldsymbol{k}}
                                   ^v-\hbar\omega_{-\boldsymbol{q}}^\mu\right)
                                   \left(\epsilon_L-\epsilon_{\boldsymbol{k}}^{c}+\epsilon_{\boldsymbol{k}}^v
                                   \right)}\\ 
\end{equation}
where $v$ and $c$ stand for valence and conduction bands, respectively, $n(\omega_{-\boldsymbol{q}}^\mu)$ is the Bose-Einstein distribution, $\epsilon_L$ is the incoming photon energy, $D_{\text{in}}$ and $D_{\text{out}}$ are the electron-photon coupling operators for the incoming and outgoing photons, respectively, $H_D$ is the defect operator, $\Delta H_{\boldsymbol{q}\mu}$ is the electron-phonon
coupling operator and $\ket{\boldsymbol{k}b}$ represents an electron in the $\boldsymbol{k}$ state in the $b$ band. All the other equations describing the scattering processes of either holes or electrons by phonons and defects are given in the supplementary.

A double-resonant process occurs when two of the three terms of the denominators of any of the $K_\alpha$ in the above equation tend to zero. As depicted in Fig.~\ref{Fig5}e, the physical meaning of this condition implies that the following are satisfied: i) the laser excites resonantly a band to band transition; ii) one of the produced electron or hole is scattered by a phonon with a quasi-momentum and an energy that matches the electronic band states; iii) one of the produced electron or hole is scattered by a defect with a quasi-momentum that matches the electronic band states. Our goal was to identify the phonon energies of all double-resonant processes as they will contribute to the Raman spectrum. Furthermore, because we expect the resonance to be enhanced when the photon energy matches an electronic transition, we considered only processes that satisfy condition i) and searched for events that satisfy either condition ii) or iii).
In all cases, these events occur such that, after all scatterings, the electron can recombine with the hole to emit the outgoing photon. More details on the calculations are provided in the supplementary, section~S3. 

Fig.~\ref{Fig5}a,b present the simulation results of the phonon-defect process and compare them with the actual experimental spectra of the monolayer taken at $\lambda_{ex}$~=~633~nm and 532~nm excitations, respectively. The general signature of the calculated phonon-defect process is the presence of central peaks associated to near-zone-center phonons that are slightly red-shifted from that of the bulk $A^1_g$ and $A^2_g$ modes. These bulk modes are not simulated, but their positions are indicated in the figure by dotted lines. Interestingly, the simulations reveal other features nearby these central peaks. In the $A^1_g$ region, a second peak is resolved on the low-energy side of the central peak. In addition, we note the presence of a small shoulder at $\lambda_{ex}$~=~532~nm that is blue shifted relative to the central peak. Changing the excitation energy leaves the central peak largely unperturbed, but shifts the features on each side (supplementary, Fig.~S5a). The simulation of the $A^2_g$ region is characterized again by a central dominant peak elongated by a wide shoulder. The shoulder extends in the low energy side to a position even below the $B_{2g}$ peak. Changing the excitation wavelength accentuates the shoulder at lower energy and produces two peaks near $B_{2g}$ (supplementary, Fig.~S5b). 

The comparison with the experimental results highlights important similarities between theory and experiment. While the spectral overlaps of the central phonon-defect peaks and bulk-allowed modes cannot be disentangled, calculated $D_1$ and $D'_2$ modes clearly match the experimentally measured shoulders and peaks found in both the $A^1_g$ and $A^2_g$ regions. We note, however, that the experiments on the monolayer do not exhibit the calculated blue shifted shoulder near the $A^1_g$ mode, nor the broad feature between the $B_{2g}$ and $A^2_g$ modes. As discussed below, these features are seen, however, in few layers samples.

Considering the approximations made for the evaluations of the matrix elements (see supplementary), it is noteworthy to mention that the simulations can not reproduce the relative intensity of these phonon-defect modes. The simulations of the double resonances at $\lambda_{ex}$~=~532~nm can, however, be rationalized using the phonon dispersion curves as a function of quasi-momentum ($q_x$, $q_y$) shown in Fig.~\ref{Fig5}c,d and superposed on top of the momentum (electronic) histograms for the $A^1_g$ and $A^2_g$ regions, respectively. This superposition highlights, for instance, the distortion in the phonon dispersions and momentum histograms along the zigzag direction, which is due to the important crystal anisotropy of P(black). Surprisingly, the calculations also show a nesting in the resonance profile near $\Gamma$ along the zigzag direction, which is visible in the middle of the histogram as two parallel stripes and pointed doughnut-like shape near $q$ = 0. This nesting is linked to the strong asymmetry of the valence band dispersion (supplementary, Fig.~S6) and contributes to the unexpectedly narrow peaks seen in the simulated spectra. 
Hence, these second-order contributions appear similar to the $D'$ modes in graphene, except that the strong anisotropy of P(black) splits the contributions according to specific lattice directions.  

Further analyses of the diagrams in Fig.~\ref{Fig5}c,d show that the red-shifted features relative to the central peaks emerge mostly from contributions in the zigzag direction, while the blue shifted features come from the armchair direction. This is visible in the figures by considering the phonon dispersion, which decreases with $q$ in the zigzag direction and increases in the armchair direction. These characteristics impose a signature or shape for each of the $D$ modes. For instance, the feature associated to the $D_1$ modes in Fig.~\ref{Fig5}c is very likely due to contributions in the zigzag direction because this is the main direction leading to red-shifted contributions.  The strong anisotropy in the dispersion relation was in fact used in our analysis of $D$ modes to determine the character of the $D$ and $D'$ families associated to dominant contributions in the zigzag and armchair directions, respectively. Obviously, further experiments would be required to provide a more quantitative assignment for each of the $D$ modes, but this is beyond the scope of the present study. By combining this model and the observations, the origin of the $D$ modes emerges, nevertheless, as being due to second-order Raman processes involving intra-valley interactions with near-zone-center phonons and defects. The simulations show also that the frequency of the $D$ modes shifts with the excitation wavelength (supplementary Fig.~S5 and Fig.~S6), but the behavior is rather complex and reflects the anisotropic phonon shifts within the momentum histograms in armchair and zigzag directions.

The evolution of the Raman response with defect density shown in Figs.~\ref{Fig3} and \ref{Fig4} is consistent with similar simulations of graphene predicting an increase in intensity at low defect density and then a decrease when the exciton diffusion length gets much longer than the defect interaction length~\cite{Lucchese2010}. According to this model, only the bulk modes ($A^1_g$ and $A^2_g$) should survive at very large defect density, which is clearly seen in the results of Fig.~\ref{Fig4}e. Based on our simulations (see more details in the supplementary) and consistent with the numbers given by the suppliers for our P(black) samples (\hyperref[sec:method]{Methods}), we estimate a concentration of defects of about 0.005\% before degradation. Using the trilayer 2D-phosphane results after 135 min exposure to ambient conditions (Fig.~\ref{Fig4}), we also estimate that a defect concentration of $\sim$1\% is enough to observe sharp bulk-allowed Raman modes, after which broadening and intensity loss are expected. Further experiments using calibrated defect density are required to explore these various regimes.

Because of the nature of the interaction, the Raman intensity of the $D$ modes should depend on the excitation wavelength, which is also seen in Fig.~\ref{Fig4}. In the case of phonon-defect modes associated to $A^1_g$, the wavelength dependence is predicted to follow~\cite{cancado_quantifying_2011}:
\begin{align}
\frac{I(D_1+D_1')}{I(A^1_g)}\cdot E{^4_{laser}}=F_D
\label{Eq2}
\end{align}
where $I(A^1_g)$ and $I(D_1+D_1')$ are the integrated intensities of the first and second-order Raman modes in the $A^1_g$ region, respectively. $E_{laser}$ is the energy of the excitation and $F_D$ is a function with an explicit dependence over the mean distance between defects. The ratio of the Raman spectra presented in Fig.~\ref{Fig4} was fitted using a model based on random defect distribution and the explicit $E_{laser}^4$ term for the Raman process.  We show in the supplementary, Fig.~S8, that the experimental $(D_1$+$D_1')$/$A^1_g$ ratio overlaps nicely with the simulated $E_{laser}^4$ dependency. This provides yet another support in favour of a phonon-defect scattering mechanism for D modes. Another observation is about the intensity of the D modes compared to first order modes, which is strong here, similar to the D modes in graphene. This behaviour contrasts with other D modes in dichalcogenides~\cite{Carvalho2013,Mignuzzi2015,Shi2016,McCreary2017}, which appear much weaker at the defect density considered here.    

The peculiar behavior of the $A^1_g$/$A^2_g$ ratio at $\lambda_{ex}$~=~532~nm with degradation observed in Fig.~\ref{Fig3} and \ref{Fig4} and reported elsewhere~\cite{favron_photooxidation_2015} can now be rationalized using the phonon-defect model. For example, the apparent blue shifted $A^1_g$ peak in Fig.~\ref{Fig3}b is associated with a change in intensity of the overlapping $A^1_g$ and $D_1$, while the broadening of the $A^2_g$ mode (supplementary, Fig.~S2) is associated with a relative increase of the $D_2'$ mode with defect density. Because of the peak degeneracy in Fig.~\ref{Fig3}, the interplay between bulk-modes and $D$ modes is difficult to resolve, but this could be addressed by considering the results in Fig.~\ref{Fig4} at different excitation wavelengths. At $\lambda_{ex}$~=~633~nm, the spectra are dominated by $D$ modes, and hence the $A^1_g$/$A^2_g$ ratio remains constant with oxidation time because of a loss of the $D$-mode intensities in both the $A^1_g$ and $A^2_g$ regions. The ratio grows only when the intensity of the bulk-allowed modes dominates due to more extensive oxidation. At $\lambda_{ex}$~=~532~nm, the relative contributions of $D$ modes are similar to that of bulk modes, but with a bigger contribution of $D$ modes in the $A^1_g$ than in the $A^2_g$ regions, resulting in an overall decrease of the ratio with oxidation. The situation at $\lambda_{ex}$~=~488~nm is similar, but provides lower contributions to the total signal from $D$ modes, which explains the relatively constant ratio with oxydation. In the light of the results in Fig.~\ref{Fig4}, it becomes apparent that the situation at $\lambda_{ex}$~=~532~nm fits in a narrow window of excitation energy where defect and bulk mode intensities is well-balanced, giving stronger variations in this ratio for sample quality assessments. Based on the model, we estimate that the intensity of $D$ modes at $\lambda_{ex}$~=~532~nm represents nearly half the integrated intensity in the $A^1_g$ region of Fig.~\ref{Fig4}c,d before degradation and that it decreases much faster than the bulk-allowed $A^1_g$ and $A^2_g$ modes. This is consistent with an $A^1_g$/$A^2_g$ ratio dominated by the phonon-defect modes at low defect level and by the bulk-allowed modes at higher defect level, which favours lower ratios.

A review of the Raman studies reported so far in the literature on mono-, bi- and few-layers 2D-phosphane films prepared using different procedures, such as i) mechanical exfoliation in inert atmosphere~\cite{favron_photooxidation_2015,phaneuf-lheureux_polarization-resolved_2016,liu_phosphorene_2014,Surrente16} and ii) in ambient conditions~\cite{Dong16,Kwon16,Ling2015,wang_highly_2015}, iii) by ultrasounds in aqueous conditions~\cite{Guo15} and iv) using plasma treatments in controlled conditions~\cite{Pei16,lu_plasma-assisted_2014}, reveal contradictory results currently undermining the use of Raman spectroscopy for quantitative analysis. For example, no consensus on the frequency evolution of $A^1_g$ from P(black) to monolayer 2D-phosphane has so far emerged: some report no obvious shift~\cite{Ling2015,lu_plasma-assisted_2014} while others report a red-shift\cite{Surrente16,phaneuf-lheureux_polarization-resolved_2016}. For the bilayer, many have reported blue-shifts ~\cite{Guo15,Pei16,phaneuf-lheureux_polarization-resolved_2016,favron_photooxidation_2015}, while some have reported a red-shift~\cite{liu_phosphorene_2014}. We believe that the influence of the newly characterized $D$ modes with defect density in the $A^1_g$ region explains these discrepancies. It is interesting to note that red-shited $A^1_g$ in both the monolayer and bilayer measured at $\lambda_{ex}$~=~532~nm were obtained from high-quality samples prepared in environmentally-controlled conditions~\cite{phaneuf-lheureux_polarization-resolved_2016,liu_phosphorene_2014,Surrente16}. As shown in Fig \ref{Fig1}b, $D_1$, located below $A_g^1$ in the monolayer, dominates the Raman spectrum at 532 nm.  However, slight level of degradation induces a fast extinction of $D_1$, which blue shifts of the emission towards that of $A_g^1$. Since $D_1$ appears to be present in the highest quality samples, this defect-activated mode can be used as an indicator of sample quality. The data shown in Fig.~\ref{Fig2}b indicates that $A_g^1$ blue-shifts from the bulk to the monolayer.

Being less affected by the presence of $D$ modes, $A^2_g$ measured at $\lambda_{ex}$~=~532~nm may prove to be a more useful indicator of sample thickness. Nonetheless, the magnitude of the blue-shift of $A_g^2$ reported in the literature for the monolayer varies considerably. For instance, mechanical exfoliation leads to a blue-shift of about $2.0\pm0.4$ $cm^{-1}$~\cite{Dong16,Surrente16,liu_phosphorene_2014,favron_photooxidation_2015,phaneuf-lheureux_polarization-resolved_2016,Ling2015,wang_highly_2015}. Plasma- and ultrasound-based exfoliation in aqueous conditions leads to slightly larger blue-shifts: 2.2-3.6 $cm^{-1}$ and 5.0 $cm^{-1}$, respectively. Our results suggest that shorter excitation wavelengths would lead to more reliable quantitative analyses of few-layer 2D-phosphane by suppressing the influence of $D$ modes in high-quality samples.

\section{\label{sec:conclusion}Conclusions}
In conclusion, we have characterized the presence of second-order Raman peaks in the spectra of n-layer black phosphorus. Based on the trends observed with sample thickness at three different excitation wavelengths, four new DDR modes, $D_1$, $D_1'$, $D_2$ and $D_2'$, are identified in the $A^1_g$ and $A^2_g$ regions. Using degradation experiments and simulations, we associate these $D$ modes to bulk-forbidden Raman modes occurring through second-order scattering. The signatures of the $D_1$, $D_2$ and $D_1'$, $D_2'$ are due to phonon-defect scattering involving predominantly intra-valley contributions in the zigzag and the armchair directions, respectively. These assignations, along with the sensitivity of $D$ modes to defects, explain the evolution of the $A^1_g$/$A^2_g$ intensity ratio during oxidation experiments. Through the use of the information obtained from $D$ modes in Raman spectroscopy, the present study rationalizes the discrepancies of Raman peak intensities and frequencies found in the literature and opens at the same time new ways to explore the properties of exfoliated P(black), such as carrier mobility, defect density, doping levels as well as chemical reactivity and mechanical stress.

\section{\label{sec:method}Methods}

Two different sources of black phosphorus samples have been used for the study: Smart Element (99.998\% purity) and HQgraphene (99.995\% purity). Despite a lower nominal purity, the second source provides wider mono-crystal ($\sim$1 $cm^2$) and better processability to make atomically thin layers. The samples were mechanically exfoliated using an adapted scotch-tape method with polydimethylsiloxane (PDMS) stamps~\cite{Meitl2006}. They were transferred on a thermally oxidized (290 nm) silicon substrate that was previously cleaned by acid treatments and baked at $300^{\circ}C$ for one hour. The produced thin layers were identified first by optical microscopy using optical contrast measurements and then measured by AFM (AutoProbe CP ThermoMicroscope) to determine the thicknesses before transferring to a Raman cryostat. The procedures were carried out at all times in a purified nitrogen glovebox.

The Raman measurements were performed in a custom micro-Raman set-up at $\lambda_{ex}$~= 488 nm, 532 nm and 633 nm laser excitations. The laser was focused on the sample with a 50~x (NA 0.5) objective with a resolution approaching the diffraction limit and a fluence ranging between 200 and 500 $\mu W\cdot\mu m^{-2}$. The spectra were acquired using a nitrogen cooled charged-coupled device camera (JY Symphony) mounted on a Jobin-Yvon Triax iHR550 spectrometer (grating 1,800 $g\cdot$$mm^{-1}$ blazed at 630 nm) with a precision of 0.2 $cm^{-1}$. During laser exposure, the sample was maintained in a homemade cell coupled to a vacuum pump providing a residual pressure of $<$  $10^{-5}$ Torr. All the spectra were calibrated in energy using the silicon peak at 520 $cm^{-1}$.

All material properties were calculated in the framework of density functional theory (DFT) with a plane wave basis as implemented in the ABINIT code~\cite{Gonze2016}. We used the PBE functional~\cite{perdew_generalized_1996} with a van-der-Waals (vdW) correction~\cite{Grimme2006} and a plane wave cutoff energy of 30 Ha. The vdW correction is needed to get an accurate spacing between the layers in bulk P(black) calculations and was kept in the monolayer calculations for consistency. In order to converge the electronic density, we used \textbf{k}-point grids of $12\times 12\times 1$ and $12\times 12\times 12$ for the monolayer and the bulk, respectively. The phonon dispersion was calculated using density functional perturbation theory~\cite{Gonze1997,Gonze2_1997} with the same converged parameters. The phonon eigenenergies were computed on $12\times 12\times 1$ and $6\times 6\times 6$ \textbf{q}-point grids for the monolayer and bulk, respectively. For the double resonant Raman intensity calculations, band structures for both electrons and phonons required much finer grids. The converged electronic density was used to calculate the electronic band structure on a fine $1000\times 1000\times 1$ \textbf{k}-point grid. The monolayer phonon band structure was interpolated on a fine $500\times 500\times 1$ \textbf{q}-point grid. For the sake of comparisons with experimental results, a scissor shift was applied to the electronic band structure to force the band gap to be the same as seen experimentally.\cite{Li2016} Also, for the phonon band structure, a scale factor was applied to get the same experimental phonon eigenenergies at $\Gamma$ as seen in the Raman spectrum. Details of the electronic structure results are given in the Supporting Information. 

\section{\label{sec:thanks}Acknowledgements} 
The authors thank E. Gaufres, D. Cardinal, and P. L. L\'evesque for technical assistance and for insightful discussions. This work was made possible by financial support from the Natural Sciences and Engineering Research Council of Canada (NSERC), the Canada Research Chairs program, the Institut de l'\'Energie Trottier, and the Fonds de Recherche du Qu\'ebec - Nature et Technologie (FRQNT). Computations were made on the supercomputer Briar\'ee from Universit\'e de Montr\'eal, managed by Calcul Qu\'ebec and Compute Canada. The operation of this supercomputer is funded by the Canada Foundation for Innovation (CFI), the minist\`ere de l'\'Economie, de la science et de l'innovation du Qu\'ebec (MESI) and the Fonds de recherche du Qu\'ebec - Nature et technologies (FRQNT).



\printbibliography
\end{document}


\maketitle

\section*{Contents}
Additional results in support to the main text are presented in this supporting information. 
The model used to compute the Raman spectra is also developed in greater
details.

  
  
  
  
  

\setcounter{figure}{0}                       
\renewcommand\thefigure{S\arabic{figure}}  

\section{\label{sec:pola}Raw Data and Polarization Experiments}
\begin{figure*}[!hbp]
\centering
 \includegraphics[trim={0cm 0cm 0cm 0cm},clip,width=16cm]{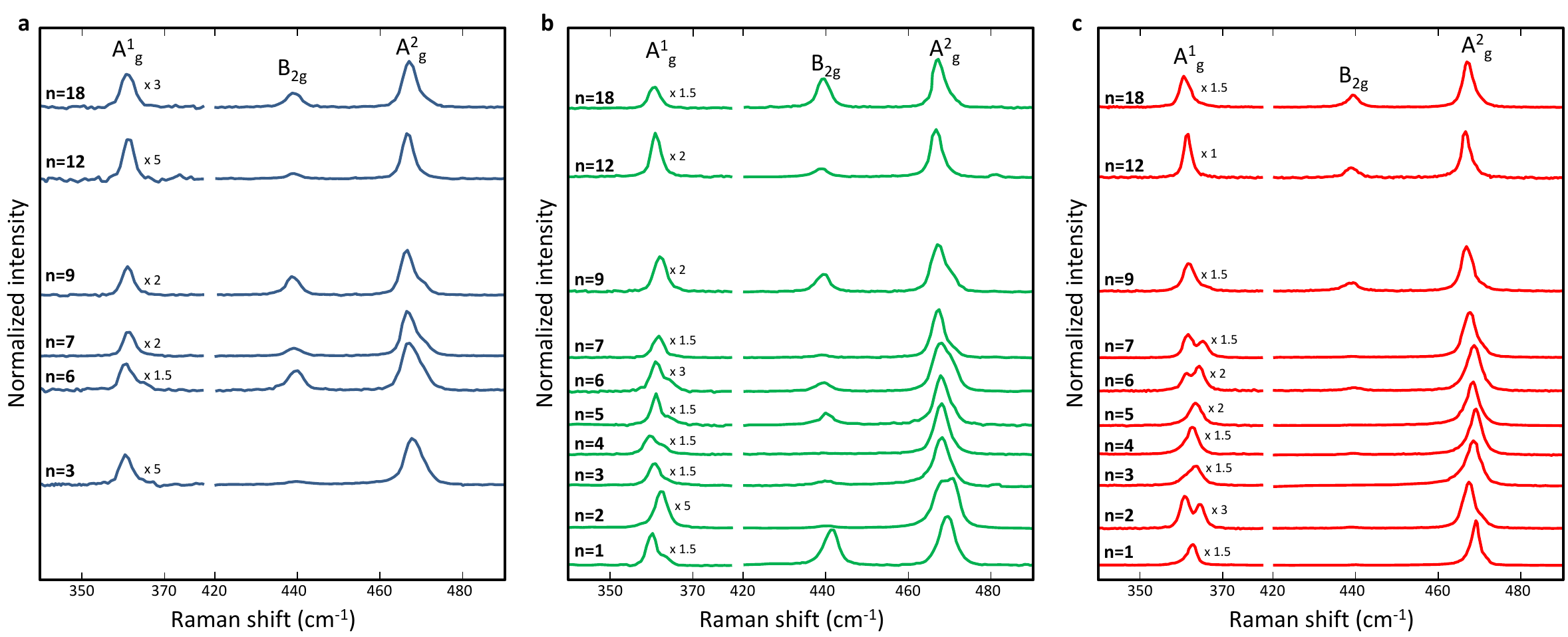}
        \caption[Raman experiments on exfoliated 2D-phosphane WITH $N$~=~1-7, 9,
        12 and 18 layers.]
        {Raman experiments on exfoliated 2D-phosphane with $n$~=~1-7, 9, 12 and 18 layers 
        deposited on a $SiO_2$/$Si$ substrate at 300 K. Raman spectra at $\lambda_{ex}$~=~488~nm ($\bf{a}$), 532~nm ($\bf{b}$) and 633~nm ($\bf{c}$) . For better clarity, the spectra are normalized relative to the $A^2_g$ maximum, vertically shifted and scaled to enhance the weaker $A^1_g$ regions. Note that a calibrated energy offset (less than 1~cm$^{-1}$) is applied to the monolayer spectrum at $\lambda_{ex}$~=~633~nm (recorded at 77~K) to match the Raman shift expected at 300~K.}
 \label{FigS1}
  \end{figure*}  
  
\begin{figure*}[!hbp]
\centering
 \includegraphics[trim={0cm 0cm 0cm 0cm},clip,width=8cm]{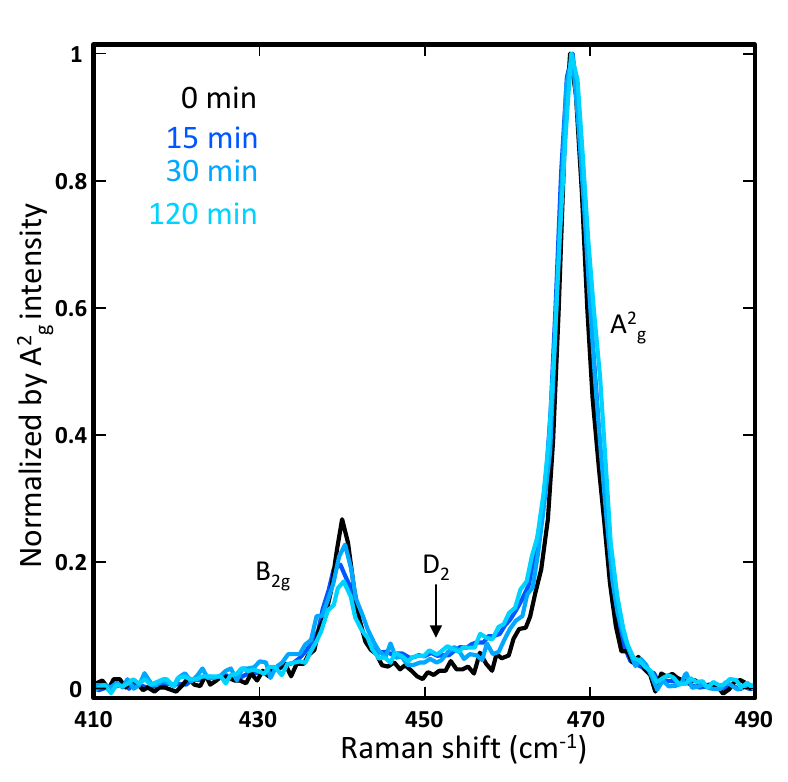}
        \caption[Raman spectra of a $n$~=~5 layer 2D-phosphane presented in
        Fig.~3 after an exposure time of 0, 15, 30 and 120 minutes to
        ambient conditions at T~=~300~K.]
        {Raman spectra of a $n$~=~5 layer 2D-phosphane presented in
        Fig.~3 after 
        an exposure time of 0, 15, 30 and 120 minutes to ambient conditions at T = 300 K.}
  \label{FigS2}
\end{figure*}

\begin{figure*}[!htbp]
 \includegraphics[trim={0cm 0cm 0cm 0cm},clip,width=16cm]{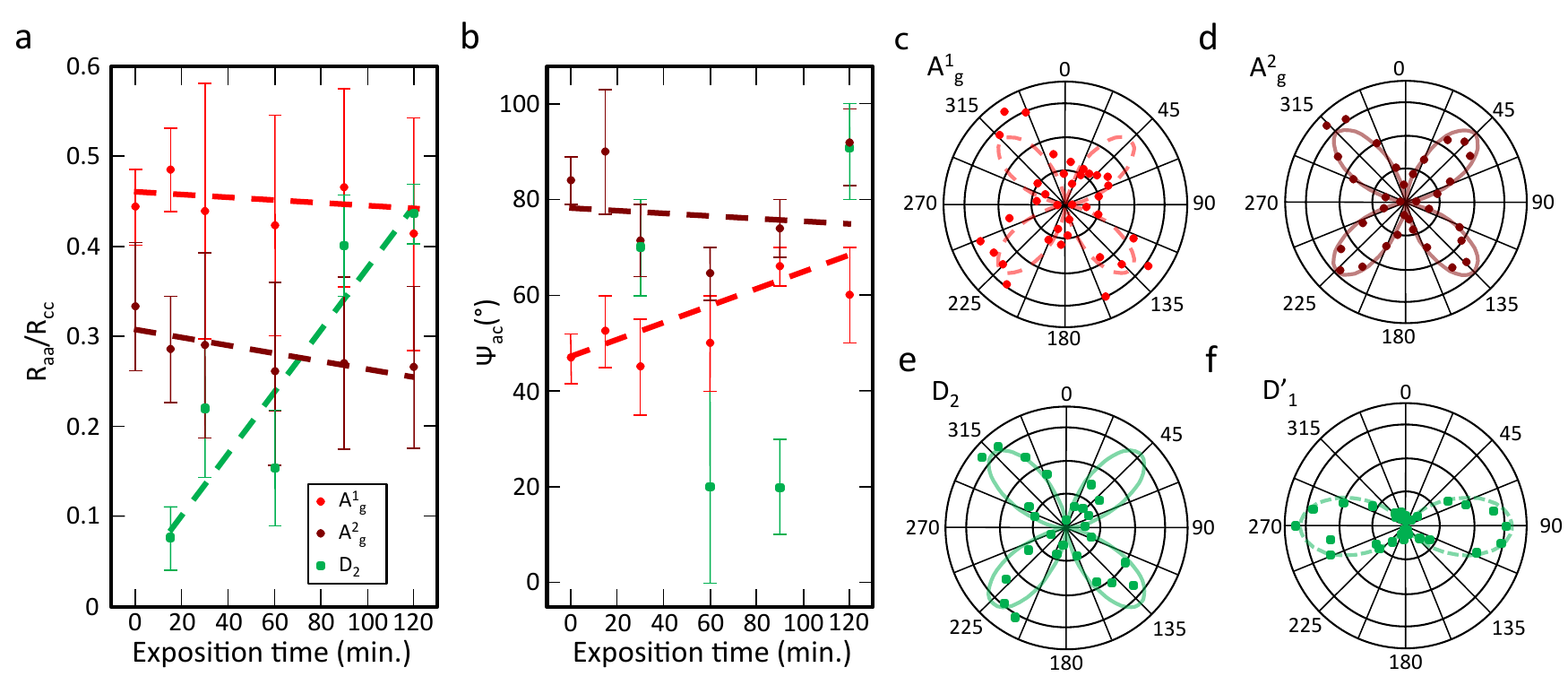}
        \caption[Perpendicular polarization measurements performed on a $n$~=~5
        layer 2D-phosphane presented in Fig.~3a-c.]
        {Perpendicular polarization measurements performed on a $n$~=~5 layer 2D-phosphane 
        presented in Fig.~3a-c, which are the perpendicular counterparts of 
        the measurements for the $A^1_g$, $A^2_g$ and $D_2$ modes, respectively, as 
        presented in Fig.~3d. The parallel polarization, noted $D_1'$ in $f$, 
        comes from a trilayer at $\lambda_{ex}$~=~532~nm and T = 300 K, 
        with no exposition to ambient condition.}
  \label{FigS3}
\end{figure*}

The Fig.~\ref{FigS1} presents the Raman spectra of all of the exfoliated black phosphorus samples at 300~K taken at three excitation wavelengths in the polarization configuration which maximizes the $A_g$ signals ($\theta_{excitation}$ = $\theta_{measured}$). Fig.~\ref{FigS2} presents the spectra of a 5-layer 2D-phosphane, presented in Fig.~3, in the $A^1_g$ region near 468 $cm^{-1}$ after multiple consecutive exposures to ambient atmosphere and illumination. The $D_2$ modes increases between 0 and 15 minutes. A small broadening of the $A^2_g$ mode is also observed, which is associated to contributions from the $D_2'$ mode.

The evolution of the fitted Raman tensor elements of $A^1_g$, $A^2_g$ and $D_2$ as a function of cumulative degradation is presented in Fig.~\ref{FigS3}.  The $R_{aa}$ and $R_{cc}$ represent the diagonal elements in the zigzag and armchair directions, respectively. $\psi_{ac}$ is the relative phase between the $a$ and $c$ directions. Information on the simulation details are available in Ref.~\cite{phaneuf-lheureux_polarization-resolved_2016}. For both the $A^1_g$ and $A^2_g$ modes, the $R_{aa}$/$R_{cc}$ ratios decrease with degradation due to a loss in anisotropy. In contrast, the ratio is significantly lower for the $D_2$ mode after 20~min exposure in ambient conditions ($R_{aa}$/$R_{cc}$ $\sim$0.1) and increases by a factor 5 after 120~min degradation. The Fig.\ref{FigS3}b presents the relative phase $\psi_{ac}$, but no trend for $D_2$ can be established. The Fig.~\ref{FigS3}c-e presents the polarization measurements in perpendicular polarization ($\theta_{excitation}$ = $\theta_{measured}$+90), which complement the polarization presented in Fig.~3d, main text. The figure also presents the polarization measurements on a $n$~=~3 layer 2D-phosphane and focuses on the $D_1'$ mode for which simulations were performed. The $R_{aa}$/$R_{cc}$ were fitted to 0.32$\pm$0.04 and $\psi_{ac}$ to 119$\pm$17$^{\circ}$. The tensor ratio (not shown) is very close to that of $A^1_g$, but with a slightly high relative phase ($R_{aa}$/$R_{cc}$=0.45$\pm$0.09 and $\psi_{ac}$= 91$\pm$12$^{\circ}$).

\section{\label{Electronic structure calculations}Electronic structure calculations}

The relaxed lattice parameters obtained are: for the bulk case along the zigzag atom arrangement, $a_1=\qty{3.324}{\angstrom}$, along the armchair atom arrangement, $a_2=\qty{4.425}{\angstrom}$ and perpendicular to the layers, $a_3=\qty{10.470}{\angstrom}$, and for the monolayer case, $a_1=\qty{3.306}{\angstrom}$ and $a_2=\qty{4.570}{\angstrom}$. These values are less than 1\% away from previous calculations and experimental values.\cite{Feng2015}    

\begin{figure*}[!t]
\centering
\includegraphics[trim={0cm 0cm 0cm 0cm},clip,width=16cm]{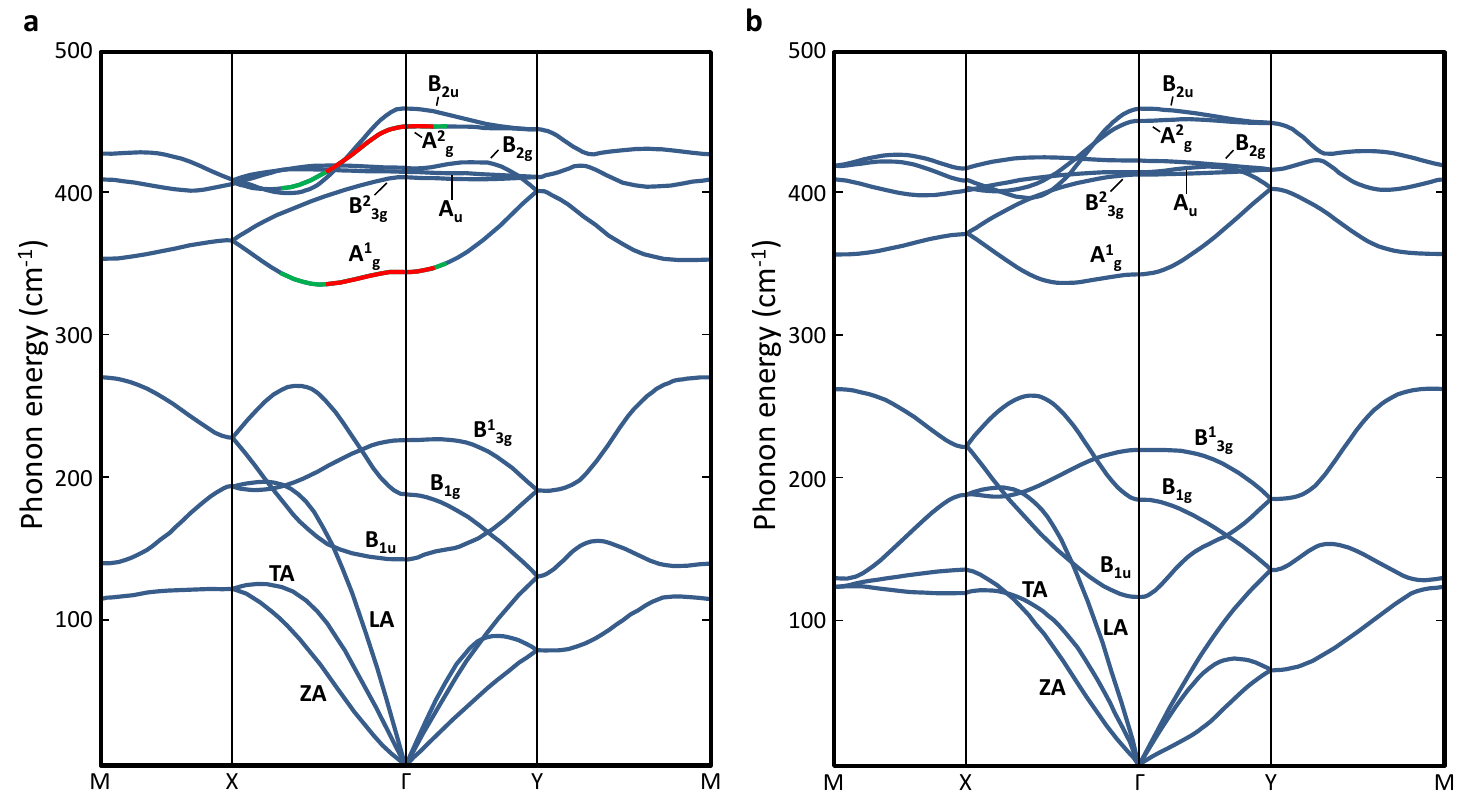}
  \caption[Phonon energy calculated for the monolayer 2D-Phosphane and bulk P(black)]
  {$\bf{a,b}$, Phonon energy calculated respectively for the monolayer 2D-Phosphane and bulk
  bulk P(black). The red and green curves in $\bf{a}$ represent the phonon range that is 
  involved in the phonon-defect process simulated for an excitation of 
  $\lambda_{ex}$~=~633~nm and 532~nm, respectively.}
  \label{FigS4}
\end{figure*}

Fig.~\ref{FigS4}\textbf{a} and Fig.~\ref{FigS4}\textbf{b} shows the phonon band structures on a conventional path in reciprocal space for the monolayer P(black) and bulk P(black), respectively, where branches labels are shown and interesting regions highlighted in green and red. Few differences are noted between both diagrams from what is seen in the figures. Since the monolayer band structure should approach the bulk band structure when adding layers, this gives a good idea of the phonon band structure for any number of layers.

The calculated phonon energies at $\Gamma$ for the monolayer is 352.6~cm$^{-1}$ for the $A_g^1$ mode and 455.9~cm$^{-1}$ for the $A_g^1=2$ mode which are within 3\% of the measured values. Note that the phonon energies have been scaled to the experimental values to ease the comparison. Furthermore, the electronic band gap which is known to be underestimated with the functional used has been corrected to the experimental value to ensure that the appropriate electronic states involved in the transition are considered in the calculations.   



\section{\label{subsec:model2}Phonon-defect model}


As mentioned in the main text, we used equation~(1) to compute the double resonant Raman spectra:
\begin{equation*}
I_{\text{DR}}(\epsilon)\propto\sum_{\boldsymbol{q}, \mu}\left|
    \sum_{\boldsymbol{k},\alpha, v, c}K_\alpha(\boldsymbol{k}, \boldsymbol{q}, \mu)\right|^2
    \delta(\epsilon_L-\epsilon - \hbar\omega^\mu_{-\boldsymbol{q}})[n(\omega_{-\boldsymbol{q}}^\mu)+1]
\end{equation*}
where all the considered $K_\alpha$ terms are listed below:
\begin{align}
\label{eq:Kee}
K_{ee}(\boldsymbol{k},\boldsymbol{q},\mu) =& \frac{\braket{\boldsymbol{k},v|D_{\text{out}}|\boldsymbol{k},c}
                                    \braket{\boldsymbol{k},c|H_D|\boldsymbol{k}+\boldsymbol{q},c}
                                    \braket{\boldsymbol{k}+\boldsymbol{q},c|\Delta H_{\boldsymbol{q}\mu}|\boldsymbol{k},c}
                                    \braket{\boldsymbol{k},c|D_{\text{in}}|\boldsymbol{k},v}}
                                   {\left(\epsilon_L-\epsilon_{\boldsymbol{k}}^{c}+\epsilon_{\boldsymbol{k}}^v
                                   -\hbar\omega_{-\boldsymbol{q}}^\mu\right)
                                   \left(\epsilon_L-\epsilon_{\boldsymbol{k}+\boldsymbol{q}}^{c}+\epsilon_{\boldsymbol{k}}
                                   ^v-\hbar\omega_{-\boldsymbol{q}}^\mu\right)
                                   \left(\epsilon_L-\epsilon_{\boldsymbol{k}}^{c}+\epsilon_{\boldsymbol{k}}^v
                                   \right)}\\
\label{eq:Keh_supplementary}
K_{eh}(\boldsymbol{k},\boldsymbol{q},\mu) =& \frac{\braket{\boldsymbol{k},v|H_D|\boldsymbol{k}+\boldsymbol{q},v}
                                    \braket{\boldsymbol{k} + \boldsymbol{q},v|D_{\text{out}}|\boldsymbol{k} + \boldsymbol{q},c}
                                    \braket{\boldsymbol{k}+\boldsymbol{q},c|\Delta H_{\boldsymbol{q}\mu}|\boldsymbol{k},c}
                                    \braket{\boldsymbol{k},c|D_{\text{in}}|\boldsymbol{k},v}}
                                   {\left(\epsilon_L-\epsilon_{\boldsymbol{k}+\boldsymbol{q}}^{c}+\epsilon_{\boldsymbol{k}+\boldsymbol{q}}^v
                                   -\hbar\omega_{-\boldsymbol{q}}^\mu\right)
                                   \left(\epsilon_L-\epsilon_{\boldsymbol{k}+\boldsymbol{q}}^{c}+\epsilon_{\boldsymbol{k}}
                                   ^v-\hbar\omega_{-\boldsymbol{q}}^\mu\right)
                                   \left(\epsilon_L-\epsilon_{\boldsymbol{k}}^{c}+\epsilon_{\boldsymbol{k}}^v
                                   \right)}\\ 
\label{eq:Khe}                                   
K_{he}(\boldsymbol{k},\boldsymbol{q},\mu) =& \frac{\braket{\boldsymbol{k},v|\Delta H_{\boldsymbol{q}\mu}|\boldsymbol{k}-\boldsymbol{q},v}
                                    \braket{\boldsymbol{k} - \boldsymbol{q},v|D_{\text{out}}|\boldsymbol{k} - \boldsymbol{q},c}
                                    \braket{\boldsymbol{k}-\boldsymbol{q},c|H_D|\boldsymbol{k},c}
                                    \braket{\boldsymbol{k},c|D_{\text{in}}|\boldsymbol{k},v}}
                                   {\left(\epsilon_L-\epsilon_{\boldsymbol{k}-\boldsymbol{q}}^{c}+\epsilon_{\boldsymbol{k}-\boldsymbol{q}}^v
                                   -\hbar\omega_{-\boldsymbol{q}}^\mu\right)
                                   \left(\epsilon_L-\epsilon_{\boldsymbol{k}}^{c}+\epsilon_{\boldsymbol{k}-\boldsymbol{q}}
                                   ^v-\hbar\omega_{-\boldsymbol{q}}^\mu\right)
                                   \left(\epsilon_L-\epsilon_{\boldsymbol{k}}^{c}+\epsilon_{\boldsymbol{k}}^v
                                   \right)}\\  
\label{eq:Khh}
K_{hh}(\boldsymbol{k},\boldsymbol{q},\mu) =& \frac{\braket{\boldsymbol{k},c|D_{\text{in}}|\boldsymbol{k},v}
                                    \braket{\boldsymbol{k},v|\Delta H_{\boldsymbol{q}\mu}|\boldsymbol{k}-\boldsymbol{q},v}
                                    \braket{\boldsymbol{k}-\boldsymbol{q},v|H_D|\boldsymbol{k},v}
                                    \braket{\boldsymbol{k},v|D_{\text{out}}|\boldsymbol{k},c}}
                                   {\left(\epsilon_L-\epsilon_{\boldsymbol{k}}^{c}+\epsilon_{\boldsymbol{k}}^v
                                   -\hbar\omega_{-\boldsymbol{q}}^\mu\right)
                                   \left(\epsilon_L-\epsilon_{\boldsymbol{k}}^{c}+\epsilon_{\boldsymbol{k}-\boldsymbol{q}}
                                   ^v-\hbar\omega_{-\boldsymbol{q}}^\mu\right)
                                   \left(\epsilon_L-\epsilon_{\boldsymbol{k}}^{c}+\epsilon_{\boldsymbol{k}}^v
                                   \right)}
\end{align}
The sum on $\alpha$ is done over all possible scattering processes. All the
considered $K$ are listed in equations \eqref{eq:Kee} to \eqref{eq:Khh} and each
of them represents a different scattering process. In all cases, the represented
scattering process starts with the production of an electron-hole pair by the
incoming photon. Before the electron and the hole are recombined, they can both
scatter on a phonon and/or on a defect as long as the scattering process gets
them on the same $\boldsymbol{k}$-point in the end. Equation $\eqref{eq:Kee}$
represents a process where both the electron scatters on a phonon and a defect
($ee$), equation \eqref{eq:Keh_supplementary} represents a diffusion of the electron by a phonon and the hole by a defect ($eh$), equation \eqref{eq:Khe} represents the scattering of the hole by a phonon and the electron by a defect ($he$) and, finally, equation~\eqref{eq:Khh} represents a process where the hole is scattered by both a phonon and a defect ($hh$).
It is noteworthy that, due to electron-phonon coupling and defect scattering, the electronic eigenstates yield a finite lifetime which would appear as an imaginary addition to all the terms in the denominator of equations \eqref{eq:Kee} to \eqref{eq:Khh}. But, for simplicity, they have been left out as they should only broaden the simulated peaks.

A double-resonance is defined as a particular event for which the corresponding $K_\alpha$ factor diverges because two of its three terms in the denominator being close to 0. In other words, when the difference between the total system energy (the sum of each quasiparticle's energies) and the laser energy is zero. Because of numerical discretization (see below) and to make sure not to miss any resonances, we define that a transition is resonant if this difference is below a certain threshold $a$. For all our calculations, we have set $a=0.005$ eV, which is large enough to not miss any double-resonance events and low enough to prevent an excessive broadening of the resonance line. The same threshold is used for scattering over a phonon or a defect. Moreover, one could find this threshold of 0.005eV is physically meaningful as it acts like a lifetime factor of $\sim$ 100 fs, which is the same order of magnitude of measured relaxation times in monolayered black phosphorus~\cite{Wang15lifetime, Wang16lifetime} for visible light excitations.

The calculated Raman spectrum is a histogram of all the phonon energies involved in a double-resonant process described above. We approximated all matrix elements in the numerators of the $K_\alpha$ terms to be 1 or 0 depending of whether the process was allowed due to symmetry considerations of the wavefunction.  Thus, the allowed $K_{ee}$ terms can be written as following:
\begin{equation}
\label{eq:equiv}
K_{ee}^{pd}(\bf{k},\bf{q},\nu) = 
     \begin{cases}
      1 &\quad\text{if} \left|\epsilon_L-\epsilon_{\boldsymbol{k}+\boldsymbol{q}}^{c}+\epsilon_{\boldsymbol{k}}
                         ^v-\hbar\omega_{-\boldsymbol{q}}^\nu\right|\leq a \text{ or if} \left|\epsilon_L-                                          \epsilon_{\boldsymbol{k}}^{c}+\epsilon_{\boldsymbol{k}}
                         ^v-\hbar\omega_{-\boldsymbol{q}}^\nu\right|\leq a\\
       0 &\quad\text{otherwise or if} \left|\epsilon_L-\epsilon_{\bf{k}}^{c}+\epsilon_{\bf{k}}^v
                                          \right|> a
     \end{cases}
\end{equation}
Similar rules apply to the other $K_\alpha$ terms for all considered events. The resulting energy histogram is then convolved using a Gaussian kernel with a 1.8 cm$^{-1}$ FWHM to smooth the histogram and replicate the experimental resolution.

As mentioned in the section {Method}, main text, calculations were done on discrete dense $\boldsymbol{k}$-point and $\boldsymbol{q}$-point grids.  This discretization imposed the introduced threshold factor $a$ to accommodate the numerical broadening between adjacent points in the grids. In all considered double-resonance processes, the electron-hole pair always scatters first on a phonon before scattering on a defect. If we had considered that the pair could scatter first on a defect (momentum transfer only), it was subject to an almost null scattering in $\boldsymbol{k}$-space since the particle could just scatter on a neighbor state with an energy difference of less than $a$ from the initial state.  Thus, this would increase the signal from near $\Gamma$ phonons by an arbitrarily factor for no apparent physical reasons. It was then decided not to consider those processes.

Finally, as discussed in the main text, we only considered phonon transitions using $A_g^1$ and $A_g^2$ branches. The $A_g^2$ phonon dispersion had to be corrected manually to preserve the band character throughout the whole Brillouin zone because the calculated eigenvalues are given by the software in ascending order, preventing any crossing between branches. Also, electronic transitions to the second conduction valley
in the armchair direction are forbidden due to the symmetries of the wavefunction (see main text). To prevent such transitions to occur, we applied a mask such that electrons do not scatter far from the first resonances.

\begin{figure*}[!htp]
\centering

 \includegraphics[trim={0cm 0cm 0cm 0cm},clip,width=16cm]{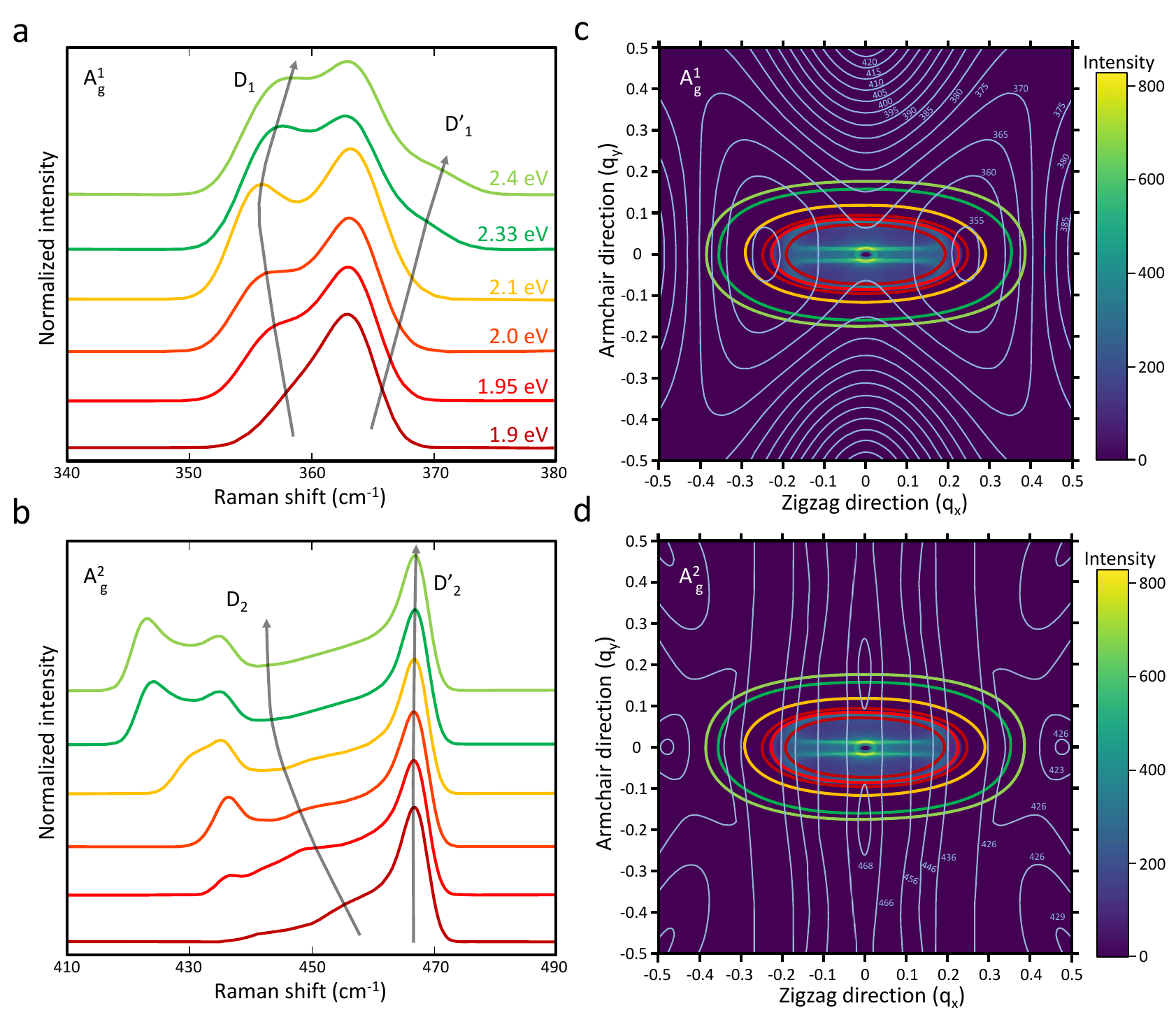}
        \caption[Simulations of the phonon-defect process.]
        {Simulations of the phonon-defect process. $\bf{a,b}$, 
        Simulated Raman spectra of phonon-defect modes for a monolayer excited at energies between 1.9 and 2.4 eV in the $A^1_g$ and $A^2_g$ mode regions, respectively. $\bf{c,d}$, Simulated momentum histogram of the phonons involved in the phonon-defect process at $\lambda_{ex}$~=~633~nm in the mode $A^1_g$ and $A^2_g$ mode regions, respectively. For comparison, the colored line show the region that can participate to the momentum histograms are also shown for excitation energies between 1.9 and 2.4 eV.}
 \label{FigSS1}
  \end{figure*}

\begin{figure*}[!hbp]
\centering
 \includegraphics[trim={0cm 0cm 0cm 0cm},clip,width=17cm]{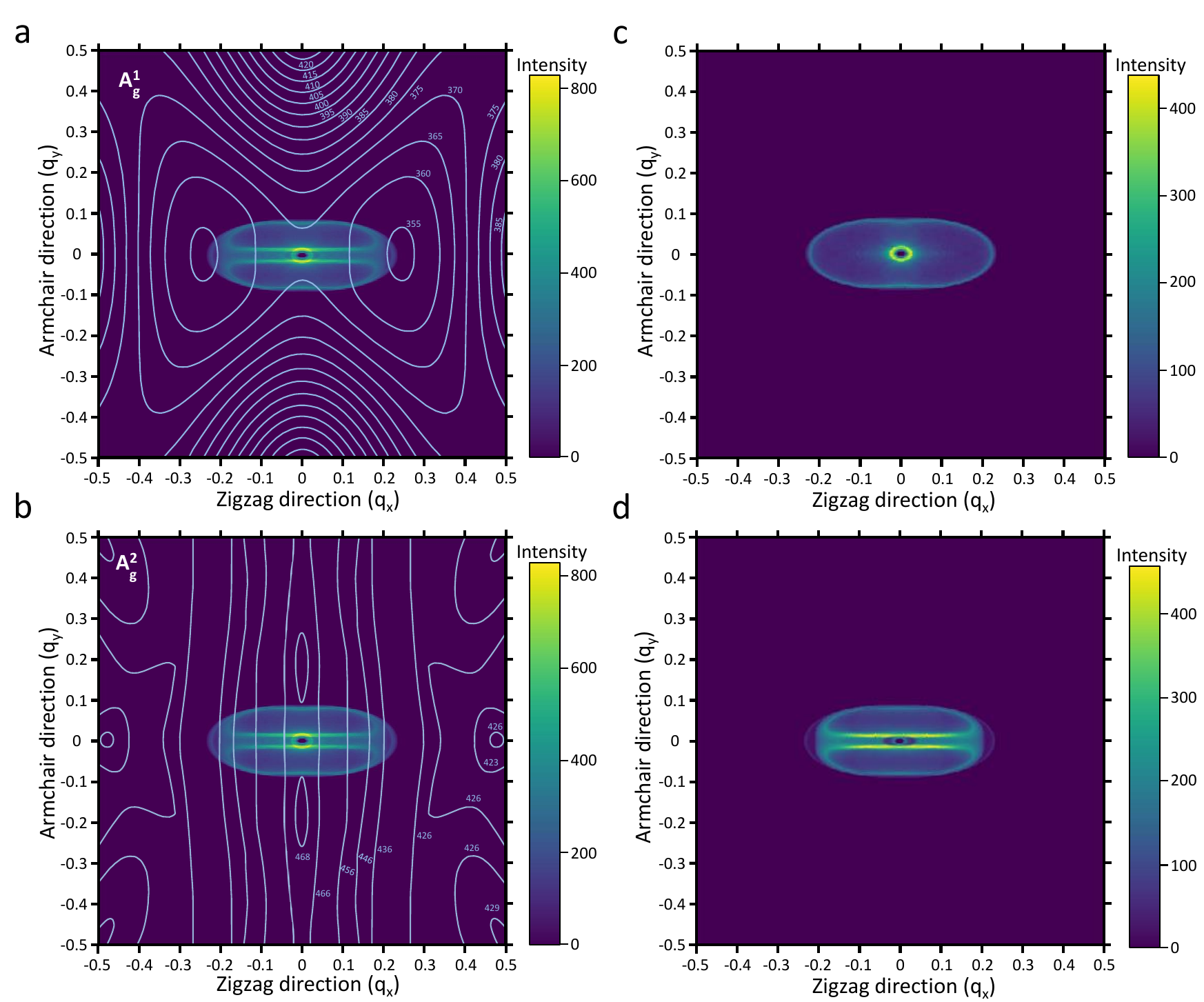}
        \caption[Simulation details of the phonon-defect process.]
        {Simulation details of the phonon-defect process. $\bf{a,b}$, Simulated momentum 
        histograms of the phonon involved in the phonon-defect process at $\lambda_{ex}$~=~633~nm in the $A^1_g$ and $A^2_g$ mode regions, respectively. $\bf{c,d}$, Momentum histograms of the phonons distributed following the contribution of the path where the phonons are emitted in the conduction band and in the valence band, respectively. A nesting is clearly seen in $\bf{a,b,d}$ in the middle of the histograms as two narrow stripes in the zigzag direction.}
 \label{FigSS2}
  \end{figure*} 
\newpage

\section{\label{sec:level1}Double-phonon simulations} 
A model based on the scattering of two phonons of opposite momenta, $q$, was investigated for the monolayer using equation~(1), main text, with the same assumptions than for the phonon-defect mode. As possible candidates, we retain every combinations that can be considered as plausible prospects due to the total energy of the sum compared to the $A^1_g$ and $A^2_g$ energies at some point in the Brillouin zone. 
 
 \begin{figure*}
 \centering
 \includegraphics[trim={0cm 0cm 0cm 0cm},clip,width=17cm]{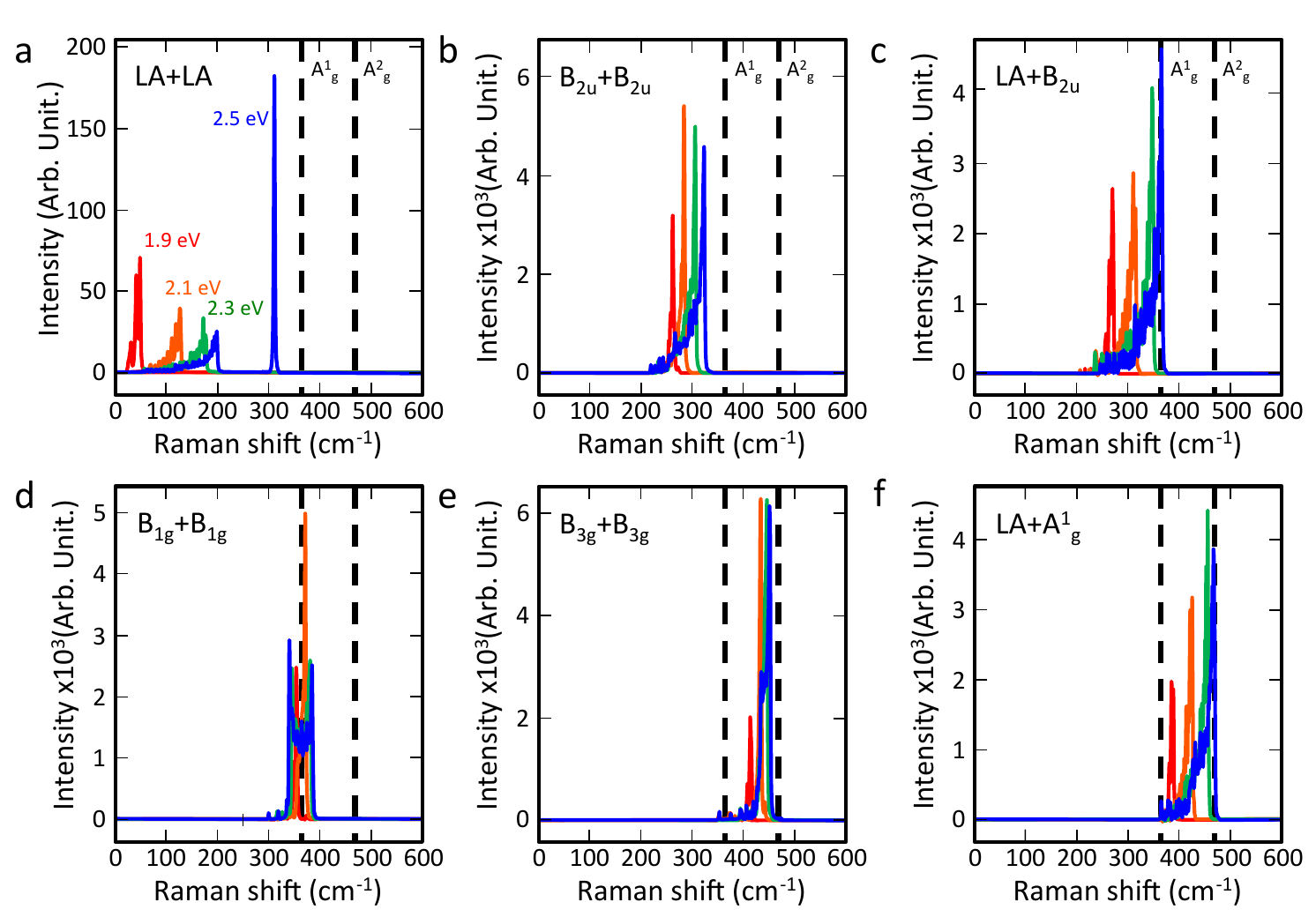}
         \caption[Simulation of different combinations of double-phonon
         resonances at four excitation energies for $n$~=~1 layer 2D-phosphane.]
         {Simulation of different combinations of double-phonon resonances at four 
         excitation energies (1.9, 2.1 2.3 and 2.5 eV) for $n$~=~1 layer 2D-Phosphane. $\bf{a-d}$, The combination of double-phonon candidates expected to appear around 363 $cm^{-1}$: LA+LA, LA+$B_{2u}$, $B_{2u}$+$B_{2u}$ and B$_{1g}$+B$_{1g}$. $\bf{e-f}$, The two combinations of double-phonon candidates expected to appear around 468 $cm^{-1}$: B$_{3g}$+B$_{3g}$ and $A^1_g$+LA. The dash line marks the position of the $A^1_g$ and $A^2_g$ modes.}
  \label{FigS5}
\end{figure*}

In addition, we consider for the double resonance only two-phonons having specific momenta that could contribute to the Raman process. The momenta that are not accessible, which are delimited in the Fig.~5c,d for the $n$~=~1 layer 2D-phosphane are avoided. Based on the polarization experiments, the final transition should have the $A_g$ symmetry. With this in hand, only four possibilities remain in the 363 $cm^{-1}$ region ($LA$+$LA$, $LA$+$B_{2u}$, $B_{2u}$+$B_{2u}$ and $B_{1g}$+$B_{1g}$) and two in the 468 $cm^{-1}$ region ($B_{3g}$+$B_{3g}$ and $A^1_g$+$LA$).

The simulation over these six candidates of two-phonon modes for $n$~=~1 layer 2D-phosphane is presented in Fig.~\ref{FigS5} for different excitation wavelengths. 
In general, the FWHMs are too high and the simulated frequencies are completely off. In addition, the simulation predicts a large shift of the peaks between $\lambda_{ex}$~=~532~nm and 633~nm, which is not observed in our experiments.
The two-phonon mode simulated cannot, however, produce sharp modes with Raman shift comparable to $A^1_g$ or $A^2_g$ modes, nor that they can they the trend seen with excitation energy.

\section{\label{defect:phonon}Defect:phonon ratio}
There is no calibration in our experiments of the density of defects with exposure time, but this number can be inferred in the limits of the experiments performed on the sample degradation. We first considered similar experiments on graphene that have shown that the Raman signatures (\emph{e.g.} the G band) is weakly affected with defects, even when the concentration reaches 1:2 (defect:C atoms)~\cite{Lucchese2010}. Although black phosphorus is more reactice~\cite{Yau1992}, it is expected that its Raman characteristics with disorder exhibit a similar behavior than graphene: \emph{i.e.} we expect a gradual increase of the FWHM of the Raman modes with defect density. As for the G band of graphene, the FWHM of the bulk modes should roughly remain constant up to a disorder level of $\sim$1\%. Interestingly, the FWHMs of the $A^1_g$ and $A^2_g$ modes remain constant in Fig.~4e. Hence, we estimate that an exposition of 135 minutes for the $n$~=~3 layer 2D-phosphane, which is the upper limit of degradation considered here, induces a defect density that is most likely lower than 1\%. Assuming a defect density proportional to exposition time, the density of defects should therefore be limited to $\sim~6\cdot10^{10}$ $cm^{-2}$ per minute.

\begin{figure*}[!hbtp]
\centering
 \includegraphics[trim={0cm 0cm 0cm 0cm},clip,width=17cm]{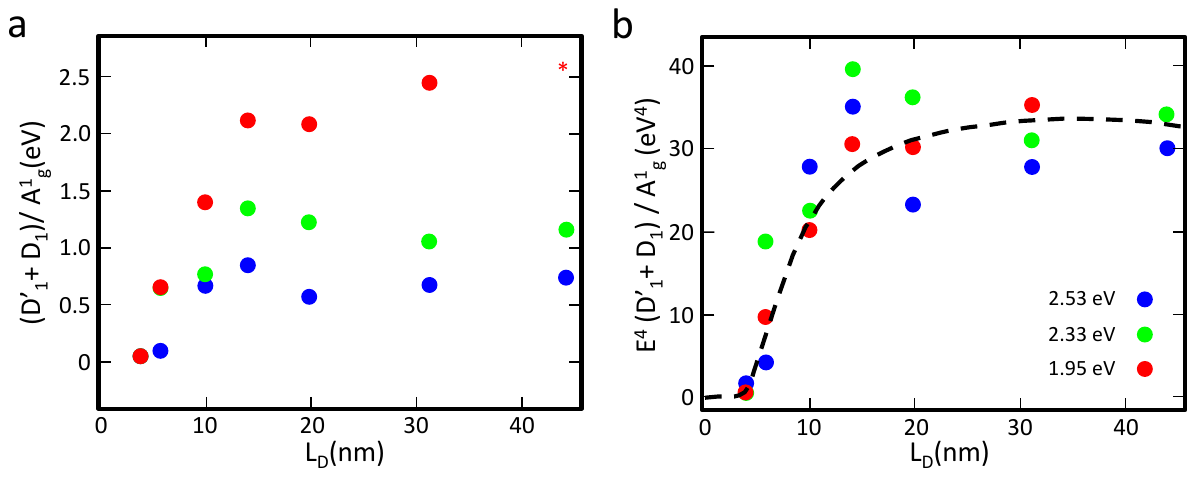}
        \caption[$D$ modes in $A_g^1$ to $A^1g$ ratio versus the estimated mean
        distance between defects.]
        {$D$ modes in $A^1_g$ to $A^1_g$ ratio versus the estimated mean distance between 
        defects (L$_D$) at $\lambda_{ex}$~=~488~nm, 532~nm and 633~nm. $\bf{a, b}$, are respectively the direct $(D_1+D_1')$/$A^1_g$ ratio and the ratio multiplied by the fourth power of the excitation energy. Note that the first Raman spectrum at $\lambda_{ex}$~=~633~nm is excluded from the analysis and added by a red star in panel $a$. The dash line in panel $b$ is a fit based on the equation in Ref.~\cite{Lucchese2010}.}
  \label{FigS6}
\end{figure*}

In the pristine state, few-layer 2D-phosphane cannot have a lower defect density than the nominal purity of the exfoliated black phosphorus, which is for some samples $\sim$~0.005 \%, giving a surface density of $\sim~3.5\cdot10^{10}~cm^{-2}$. 
While we did not know the initial degradation state of the n=3 layer 2D-phosphane, it is clear therefore that it is most likely higher than $3.5\cdot10^{10}$ $cm^{-2}$ in the initial state right after the exfoliation of the sample. Moreover, we estimate a degradation rate of $5\cdot10^{10}$ $cm^{-2}min^{-1}$ for the $n$~=~3 layer 2D-phosphane, which gives an initial density of defects of $5\cdot10^{10}$ $cm^{-2}$. The simulation in Fig.\ref{FigS6} presents the dependence of $(D_1+D_1')$/$A^1_g$ ratio for three different wavelengths with the mean distance between defect ($L_D$) based on the estimation of the density.

\printbibliography